\pdfoutput=1
\documentclass[journal=accacs,manuscript=article,layout=twocolumn]{achemso}
\usepackage[version=4]{mhchem} 

\newcommand\tabcaption{\def\@captype{table}\caption}
\newcommand\figcaption{\def\@captype{figure}\caption}
\usepackage{amsmath}
\usepackage{algorithm}
\usepackage[noend]{algpseudocode}
\usepackage{chemformula}    
\usepackage{bm}
\usepackage{multirow}
\usepackage{mdframed}
\usepackage{rotating}
\usepackage{booktabs}
\usepackage{etoolbox}
\usepackage{enumitem}
\usepackage{subcaption}
\usepackage[nonumberlist]{glossaries}
\mciteErrorOnUnknownfalse
\usepackage{xcolor}
\usepackage{comment}
\usepackage{array}
\usepackage[hidelinks]{hyperref}
\newcounter{magicrownumbers}
\preto\tabular{\setcounter{magicrownumbers}{0}}
\usepackage[textsize=small]{todonotes}
\usepackage{siunitx}
\usepackage{amssymb}
\usepackage{pifont}
\usepackage{threeparttable}
\usepackage{graphicx}
\usepackage{longtable}
\usepackage{float}
\usepackage{placeins}
\usepackage[utf8]{inputenc}

\newcommand{\dataset}[1]{ODAC23}

\newacronym{DFT}{DFT}{Density Functional Theory}
\newacronym{SI}{SI}{Supporting Information}
\newacronym{PES}{PES}{Potential Energy Surface}
\newacronym{QE}{QE}{Quantum Espresso}
\newacronym{VASP}{VASP}{Vienna Ab initio Simulation Package}
\newacronym{OC20}{OC20}{Open Catalyst 2020 Dataset}
\newacronym{ML}{ML}{machine learning}
\newacronym{GNNs}{GNNs}{Graph Neural Networks}
\newacronym{GNN}{GNN}{Graph Neural Network}
\newacronym{MD}{MD}{\textit{ab initio} Molecular Dynamics}
\newacronym{MAE}{MAE}{Mean Absolute Error}
\newacronym{ODAC23}{ODAC23}{Open DAC 2023}
\usepackage[version=4]{mhchem}

\glsaddall

\newcommand{\gatech}{School of Chemical and Biomolecular Engineering, Georgia Institute of Technology, Atlanta, GA, USA}
\newcommand{\ornl}{Oak Ridge National Laboratory, Oak Ridge, TN, USA}

\author{Logan M. Brabson}
\affiliation{\gatech}
\author{Andrew J. Medford}
\affiliation{\gatech}
\email{ajm@gatech.edu}
\author{David S. Sholl}
\affiliation{\ornl}
\email{shollds@ornl.gov}

\title[]
  {Comparing classical and machine learning force fields for modeling deformation of solid sorbents relevant for direct air capture}

\abbreviations{IR,NMR,UV}

\let\oldmaketitle\maketitle
\let\maketitle\relax

\begin{document}
\twocolumn[
\begin{@twocolumnfalse}
\oldmaketitle
\begin{abstract}
\input{insbox}
\InsertBoxR{0}{
  \vspace{-5pt}%
  \includegraphics[width=0.3\linewidth]{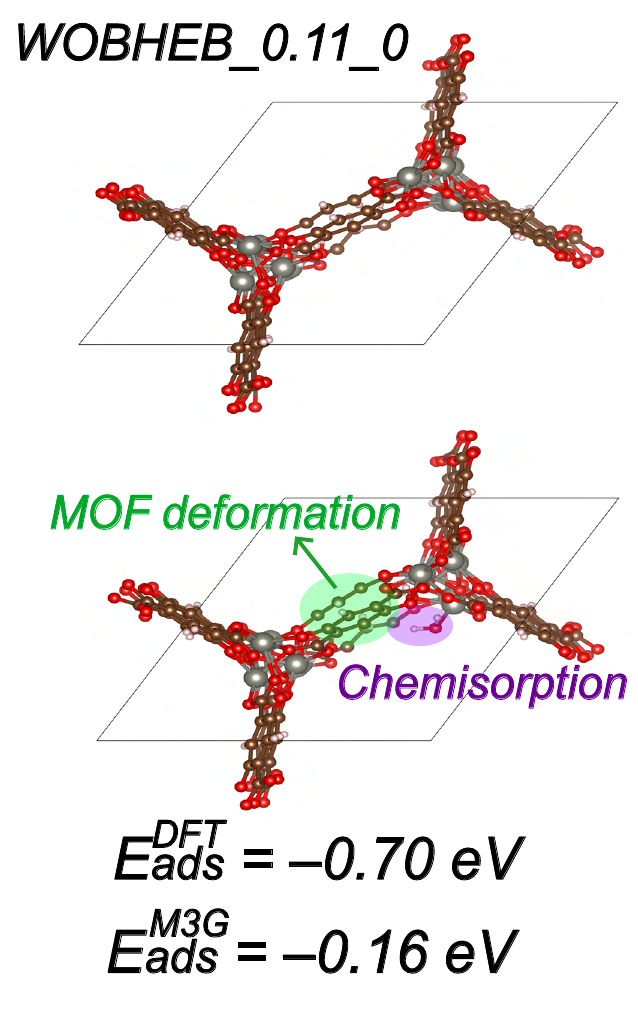}
}
Direct air capture (DAC) with solid sorbents such as metal–organic frameworks (MOFs) is a promising approach for negative carbon emissions. Computational materials screening can help identify promising materials from the vast chemical space of potential sorbents. Experiments have shown that MOF framework flexibility and deformation induced by adsorbate molecules can drastically affect adsorption properties such as capacity and selectivity. Force field (FF) models are commonly used as surrogates for more accurate density functional theory (DFT) calculations when modeling sorbents, but most studies using FFs for MOFs assume framework rigidity to simplify calculations. Although flexible FFs for MOFs have been parameterized for specific materials, the generality of FFs for reliably modeling adsorbate-induced deformation to near-DFT accuracy has not been established. This work benchmarks the efficacy of several general FFs in describing adsorbate-induced deformation for DAC against DFT. Specifically, we compare a common classical FF (UFF4MOF) with several machine learning (ML) FFs: M3GNet, CHGNet, MACE-MP-0, MACE-MPA-0, eSEN, and the Equiformer V2 model developed from the recent Open DAC 2023 dataset. Our results show that current classical methods are insufficient for describing framework deformation, especially in cases of interest for DAC where strong interactions exist between adsorbed molecules and MOF frameworks. The emerging ML methods we tested -- particularly CHGNet, MACE-MP-0, and Equiformer V2 -- appear to be more promising than the classical FF for emulating the deformation behavior described by DFT but fail to achieve the accuracy required for practical predictions.

\vspace{1.5cm}
\begin{minipage}{0.9\textwidth}
Notice of Copyright: This manuscript has been authored by UT-Battelle, LLC, under contract DE-AC05-00OR22725 with the US Department of Energy (DOE). The publisher acknowledges the US government license to provide public access under the DOE Public Access Plan (http://energy.gov/downloads/doe-public-access-plan).
\end{minipage}
\end{abstract}

\end{@twocolumnfalse}
]

\clearpage
\section{Introduction}

Direct air capture (DAC) seeks to capture large quantities of carbon dioxide directly from atmospheric air but remains a challenging scientific problem due to both the low partial pressure of \ce{CO2} in the air and the need to identify adsorbents suitable for repeated use in cost-effective processes.\cite{SevenSep, Beuttler2019, DAC_review, Kim2022} Metal–organic frameworks (MOFs) are a promising class of solid sorbents for DAC at low temperatures given their high porosities, modularity, and tunability.\cite{Furukawa2013, Moosavi2020, Deria2015} In contrast with more traditional liquid DAC sorbents such as alkali hydroxides, MOFs can bind \ce{CO2} via physisorption in addition to chemisorption and typically require less energy for regeneration.\cite{Yang2024, Tiainen2022}

One important consideration when using MOFs for gas separations is their flexibility.\cite{Senkovska2023, Agrawal2019, Han2020} Framework flexibility can be either intrinsic or induced by guest molecules. Intrinsic flexibility exists in all MOFs as a result of thermal vibrations, whereas adsorbate-induced deformation only occurs in some MOFs.\cite{Daou2021} Adsorbate-induced deformation can manifest as significant structural and unit cell volume changes, like in breathing or swelling MOFs,\cite{Serre2007,Mason2015} or as more subtle changes such as linker rotation.\cite{Jawahery2017} Accounting for flexibility makes computational models more physically accurate, and accounting for flexibility indeed has a non-negligible influence on properties of interest.\cite{Dundar2017} 

Despite flexibility playing a role in accurate modeling of adsorption, almost all large-scale computational studies of MOFs for gas separations assume MOF rigidity and only consider non-bonded interactions during molecular dynamics or Monte Carlo simulations.\cite{Daglar2020,Yang2022,Oliveira2023,Boyd2017} Yu et al. showed that molecular simulations performed with this rigidity assumption can cause significant underestimations of gas uptake for a variety of physisorbed molecules in MOFs at dilute conditions and proposed several possible strategies for approximating framework flexibility in molecular simulations.\cite{ZhenziYu2021} These and related strategies, however, suffer from either insufficient accuracy or prohibitively high computational cost for high-throughput studies.\cite{Rogge2019} Unfortunately, the ability of common models such as classical force fields to adequately describe MOFs with intrinsic flexibility\cite{Witman2017,Agrawal2019,Daou2021} or MOFs that undergo structural deformation in the presence of guest molecules (adsorbate-induced deformation)\cite{Jawahery2017, Coudert2011} is not well established. 

\begin{figure*}[ht!]
    \centering
    \includegraphics[width=\textwidth]{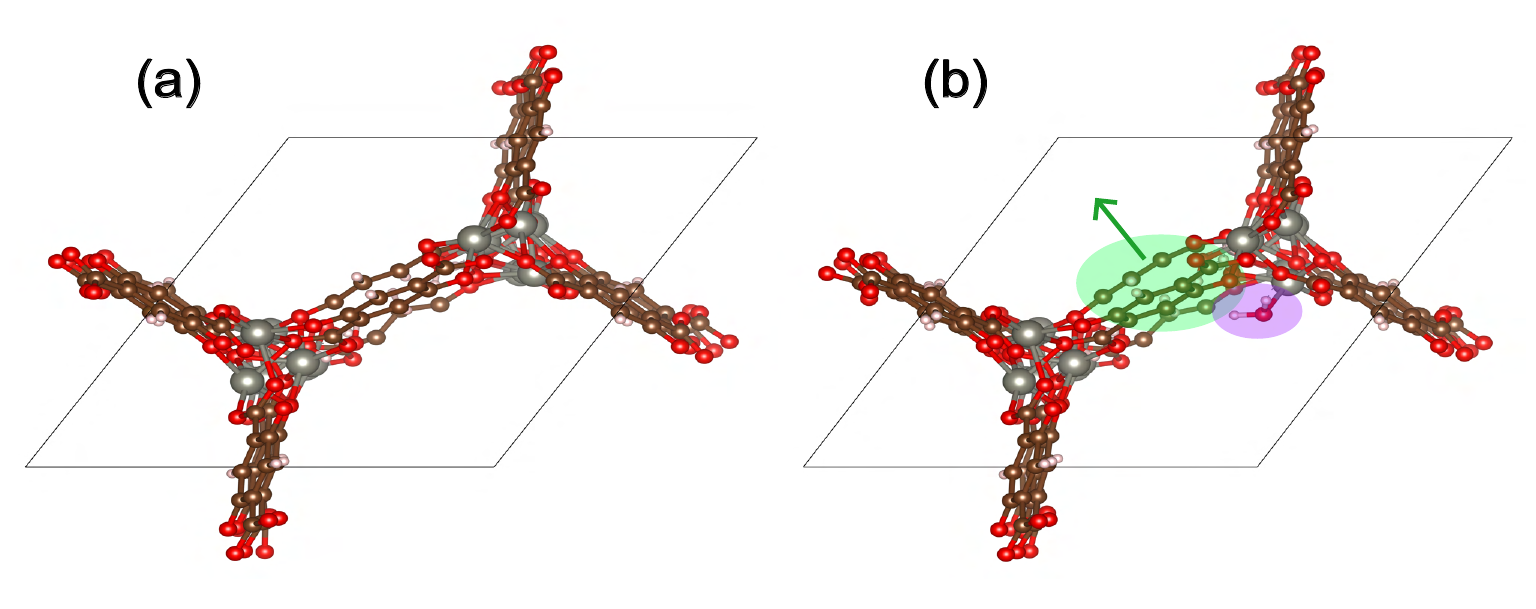}
    \caption{DFT-relaxed structures for the defective MOF with CSD code WOBHEB (system ID 0\_217) (a) before and (b) after the introduction of one \ce{H2O} adsorbate molecule, shaded in purple. The green arrow indicates the expansion of the linker upon adsorption. Zn, C, O, and H atoms are shown as silver, brown, red, and white, respectively.}
    \label{fig:deform_vis}
\end{figure*}

An example of a MOF that undergoes adsorbate-induced deformation from the large collection of DFT results from the recent Open DAC 2023 work\cite{ODAC23} can be seen in Figure \ref{fig:deform_vis}. In this MOF, the introduction of a \ce{H2O} guest molecule, shaded in purple, yields a relaxed structure with the linker, shaded in green, expanded outward relative to its position in the empty MOF. Without the \ce{H2O} molecule included, the MOF structure on the right is more than 0.3 eV/unit cell less favorable. This deformation energy is significant relative to adsorption energies that typically range from 0 to 1 eV. Although the visual changes in geometry are minor, these slight changes in atomic positions can add up over the 100+ atoms in a typical MOF unit cell.

The vast number of possible MOFs implies that computation can play an important role in screening materials, provided the underlying models have sufficient accuracy.\cite{Lee2021} The high computational cost of quantum mechanical (QM) calculations such as density functional theory (DFT) necessitates the use of force fields (FFs) to improve efficiency of high-throughput materials screening (HTS) studies.\cite{Nazarian2017,Li2022,Qiao2016,Islamov2023,Schmidt2015,Harrison2018} FFs have been developed to account for MOF flexibility, but they are typically system-specific and exhibit poor transferability.\cite{Bureekaew2013,Heinen2018} Machine learning force fields (MLFFs) trained on QM data can bridge the gap between classical FFs and ab initio methods, but they must be applicable beyond their underlying training data to be used widely.\cite{Unke2021, Ko2023} Some emerging ``foundational'' MLFFs based on deep graph neural networks seek to universally describe the entire periodic table, such as the Materials Graph Network with three-body interactions (M3GNet),\cite{MEGNet,M3GNet} Crystal Hamiltonian Graph Neural Network (CHGNet),\cite{Deng2023} and the Multi-Atomic Cluster Expansion (MACE)\cite{MACE_arch,Drautz2019} architecture.

Batatia et al. released several pre-trained MACE models called MACE-MP-0, and they showed that the (medium-sized) MACE-MP-0 MLFF was widely applicable to atomic structures for catalysis and to aqueous systems.\cite{MACE-MP-0} MACE-MP-0 performs well at predicting MOF energies for $\geq$20,000 DFT-relaxed MOFs from the Quantum MOF database\cite{Rosen2021} and at describing \ce{CO2} adsorption in Mg-MOF-74. M3GNet, CHGNet, and MACE-MP-0 were trained on the Materials Project (MP) database of inorganic crystals.\cite{Jain2013_MP} The CHGNet and MACE-MP-0 training sets are an order of magnitude larger than that of M3GNet because CHGNet and MACE-MP-0 were trained on the \texttt{MPtrj} subset of the MP database, which includes structural relaxation trajectories for the MP materials.\cite{Deng2023} While the MP database contains a large number of solids, it is biased heavily toward oxides and contains relatively few MOFs. Batatia et al. released an updated MACE model called MACE-MPA-0 trained on 3.5 million structures from the MPtrj and sAlex databases. The sAlex dataset was developed as part of the Open Materials 2024 project.\cite{alexandriaDatabase,OMAT24} MACE-MPA-0 outperforms MACE-MP-0 on Matbench benchmarks,\cite{Riebesell2023} though the authors caution that improved accuracy is not guaranteed in all cases. Most recently, Fu et al. trained their equivariant Smooth Energy Network (eSEN) on the combined OMat, MPtrj, and sAlex datasets with an emphasis on conserving energy during molecular dynamics simulations. eSEN achieved stat-of-the-art results on physical property predictions tasks including materials stability and thermal conductivity predictions.\cite{eSEN} Table \ref{tab:models} shows the size of the training dataset and the number of model parameters for the MLFFs considered in this study.

Recent work has made progress in assessing the accuracy of universal MLFFs for various chemical applications. Focassio et al. showed that three common MLFFs -- MACE-MP-0, CHGNet,\cite{Deng2023} and M3GNet -- did not adequately describe a set of 1,497 solid surfaces including 73 unique elements but that these models could be useful for fine-tuning more specialized models.\cite{Focassio2024} For MOFs, Zheng et al. showed that molecular dynamics using the Deep Potential-Smooth Edition (DeepPot-SE)\cite{Zheng2023} MLFF is more capable than UFF in describing Mg-MOF-74 flexibility, which enhances \ce{CO2} diffusivity in the pores compared to rigid models.\cite{Zheng2024} Riebesell et al. published a leaderboard for ranking MLFFs by their ability to predict MOF thermodynamic stability, but no comparison to a classical FF was provided.\cite{Riebesell2023} They found that MLFFs trained on forces and stresses from DFT relaxation trajectories, including M3GNet and MACE, outperform MLFFs that do not. Despite these advances, no systematic comparison of classical and universal MLFFs has been performed for MOFs in the context of an application such as DAC.

This work builds upon the recent Open DAC 2023 (ODAC23) dataset of more than 38 million DFT calculations for \ce{CO2} and \ce{H2O} adsorption in MOFs.\cite{ODAC23} That work sought to raise the baseline level of theory for computational DAC studies in MOFs from classical FFs (i.e., UFF) to DFT. As might be expected, the accuracy of UFF compared to DFT results differs significantly between physisorbed and chemisorbed molecules. The ODAC23 project also trained the Equiformer architecture,\cite{Liao2022_equi} previously used for the Open Catalyst project,\cite{OC20,OC22} to create a customized ML model capable of directly predicting the adsorption energy of \ce{CO2} or \ce{H2O} in MOFs. Although this model outperformed UFF, a more thorough comparison of Equiformer's performance relative to other MLFFs is needed, particularly in cases where deformation is significant.

In this paper, we use the DFT calculations in the ODAC23 dataset to compare the ability of various FFs to describe adsorbate-induced deformation in MOFs relevant for DAC. We selected a diverse set of 60 MOF+adsorbate systems from the ODAC Equiformer test set, split between systems that undergo significant adsorbate-induced deformation and those that do not. We prioritized including MOFs that are potentially favorable for DAC given their selectivity toward adsorption of \ce{CO2} over \ce{H2O} according to ODAC23 DFT calculations.\cite{Findley2021} We begin by showing that adsorption energy calculations in rigid and flexible MOFs frequently differ regardless of the method used, challenging the rigidity assumption commonly used in Monte Carlo simulations. Taking DFT adsorption energies as the ground truth, we then compare a widely used classical FF (UFF4MOF, referred to as UFF in this work for simplicity)\cite{UFF4MOF1,UFF4MOF2} to M3GNet, CHGNet, eSEN, and the pre-trained MACE-MP-0 and MACE-MPA-0 models. We also include the tailored Equiformer V2 (Large) model from ODAC23 (EqV2-ODAC) in our comparison.\cite{ODAC23}. Three of the MLFFs, namely CHGNet, MACE-MP-0, and EqV2-ODAC, outperform the classical approach in predicting adsorption energies in MOFs that undergo adsorbate-induced deformation. The relative success of CHGNet and MACE-MP-0 given their training data is surprising and deserves further study. Unfortunately, we do not identify a model that predicts a mean absolute adsorption energy error relative to DFT of less than 0.1 eV, a reasonable upper bound on errors for practical calculations.
The findings suggest that accounting for MOF flexibility is critical to achieve accurate chemisorption predictions, but that additional force field development is needed to achieve this across all MOFs.

\begin{table*}[t]
    \centering
    \renewcommand{\arraystretch}{1.0}
    \setlength{\tabcolsep}{5pt}
    \renewcommand{\arraystretch}{1.0}
    \setlength{\tabcolsep}{6pt}
    \resizebox{0.9\linewidth}{!}{
    \begin{tabular}{lrrrrr}
        \toprule
        Model & \ \# of training structures & \ \# of parameters & \ Force consistency \\
        \midrule

M3GNet\cite{M3GNet} & 188k & 228k & Yes \\
CHGNet\cite{Deng2023} & 1.58M & 413k & Yes \\
MACE-MP-0 (medium)\cite{MACE-MP-0} & 1.58M & 4.69M & Yes \\
MACE-MPA-0 (medium)\cite{MACE-MP-0} & 11.98M & 9.06M & Yes \\
eSEN-30M-OAM\cite{eSEN} & 113M & 30.2M & Yes \\
ODAC Equiformer V2 (large)\cite{ODAC23} & 38M & 153M & No \\

    \bottomrule
    \end{tabular}
    }
    \caption{Training dataset size, model size, and force consistency for the six MLFFs used in this study; force consistent models calculate forces as the gradient of the energy surface}
    \label{tab:models}
    \vspace{-10pt}
\end{table*}
\section{Methods}

\subsection{Energy definitions}
A key quantity in any description of molecular adsorption in MOFs is the adsorption energy for a single molecule, $E_{ads}$. The adsorption energy is defined by

\begin{align}
E_{ads} = E_{sys}(r_{sys}) – E_{MOF}(r_{MOF}) -  \nonumber \\
– E_{adsorbate}(r_{adsorbate}) 
\label{eq:Eads}
\end{align}

\noindent where in $E_{x}(r_{y})$, $x$ refers to the set of atoms considered in the energy calculation, with $y$ defining the atoms included in the geometry used for relaxation. The subscript $sys$ refers to a combined MOF+adsorbate geometry and the subscript $adsorbate$ refers to the isolated gas-phase adsorbate. All energy terms are at relaxed geometries.

In Figure \ref{fig:deform_vis}, $r_{MOF}$ and $r_{sys}$ refer to the geometries shown in (a) and (b), respectively. $E_{ads}$ is the primary metric of interest in this work, and a good FF for practical simulations should agree with DFT to within $\pm$0.1 eV. To highlight the influence of deformation of the MOF framework, it is useful to decompose $E_{ads}$ into energies associated with interaction ($E_{int}$) and MOF deformation ($E_{MOF,deform}$). As we will show, this decomposition is critical to deconvoluting the influence of MOF deformation, but we emphasize that the physically significant quantity is $E_{ads}$.

We define the MOF-molecule interaction energy as:

\begin{align}
E_{int} = E_{sys}(r_{sys}) – E_{MOF}(r_{sys}) \nonumber \\
– E_{adsorbate}(r_{sys}) 
\label{eq:Eint}
\end{align}

\noindent Unlike the adsorption energy, the geometry of the MOF used in every term in Eq. (\ref{eq:Eint}) is the same, so that the interaction energy captures only the electronic contributions. Notably, $E_{adsorbate}(r_{sys})$ implies that periodic boundary conditions are used, which will remove any self-interactions between the adsorbates from $E_{int}$. In practice, it is often more convenient to use non-periodic vacuum conditions for the adsorbate, which introduces an adsorbate self-interaction energy:

\begin{align}
E_{adsorbate,self} = E_{adsorbate}(r_{sys}) \nonumber \\
– E_{adsorbate}(r_{sys,vacuum})
\label{eq:Eself_ads}
\end{align}

\noindent $E_{adsorbate,self}$ quantifies the energy that arises due to placing adsorbates in a periodic system, and should be negligible in low loadings or for MOFs with large unit cells.

In a typical high-throughput molecular simulation assuming MOF rigidity, the position of framework atoms in $r_{sys}$ are identical to the crystal structure of the empty MOF, so $r_{sys} = r_{MOF}$ and $E_{int}$ becomes equivalent to $E_{ads}$ when $E_{adsorbate,self}$ can be neglected.
A more physically complete description of adsorption recognizes that the MOF geometry after adsorption is different than in the empty MOF structure, so $E_{ads}$ is not equal to $E_{int}$.

To quantify the contributions to adsorption due to MOF and adsorbate deformation, it is useful to define

\begin{align}
E_{MOF,deform} = E_{MOF}(r_{sys}) – E_{MOF}(r_{MOF})
\label{eq:Edef_MOF}
\end{align}

\begin{align}
E_{adsorbate,deform} = E_{adsorbate}(r_{sys,vacuum}) \nonumber \\
– E_{adsorbate}(r_{adsorbate})
\label{eq:Edef_ads}
\end{align}

\noindent The energies on the right-hand sides of Equations \ref{eq:Edef_MOF} and \ref{eq:Edef_ads} are for the same sets of atoms at different geometries. These energies describe the effects of atomic movements during adsorption for the MOF and adsorbate, respectively.

Equations \ref{eq:Eads}-\ref{eq:Edef_ads} can be combined to show that the difference between the adsorption and interaction energies is equivalent to the sum of the MOF deformation, the adsorbate deformation, and the adsorbate self-interaction energy:

\begin{align}
E_{ads} – E_{int} = E_{MOF,deform} + E_{adsorbate,deform} \nonumber \\ + E_{adsorbate,self}
\label{eq:Ediff}
\end{align}

\noindent Most molecular simulations of small molecules treat the molecule as rigid such that the $E_{adsorbate,deform}$ term in Eq. \ref{eq:Ediff} is exactly zero. In DFT calculations, the molecule can deform, so this term cannot necessarily be omitted. 

It is also useful to define an energy term representative of the binding energies used in molecular simulations where the MOF is held rigid. For example, consider a grand canonical Monte Carlo (GCMC) simulation to compute the uptake of a guest molecule in a MOF. The key quantity used to accept or reject trial moves in such simulations is the binding energy of the guest molecule to the framework, and almost all published GCMC simulations assume MOF rigidity. Thus, neither $E_{ads}$ nor $E_{int}$ are directly applicable to this situation because, in this work, we have relaxed the rigid MOF assumption and both $E_{ads}$ and $E_{int}$ explicitly allow for the movement of framework atoms upon adsorption. To disambiguate, we define the static adsorption energy as

\begin{align}
E_{static} = E_{sys}(r_{MOF}, r_{adsorbate}) – E_{MOF}(r_{MOF}) \nonumber \\
– E_{adsorbate}(r_{adsorbate})
\label{eq:Egcmc}
\end{align}

\noindent $E_{static}$ is analogous to $E_{ads}$ with the MOF being held rigid during combined MOF+adsorbate relaxation. That is, $r_{MOF}$ in the $E_{sys}$ term is the geometry determined from empty MOF relaxation in the $E_{MOF}$ term of Eq. \ref{eq:Egcmc}, and only the adsorbate positions were allowed to change during system relaxation. 

$E_{static}$ can be considered a type of interaction energy since the MOF is held rigid during relaxation but differs from the definition of $E_{int}$ in Eq. \ref{eq:Eint} because the MOF geometry for $E_{static}$ is determined from relaxation of the empty MOF without a guest molecule. That is, $E_{static}$ is equivalent to the adsorption ($E_{ads}$) and interaction ($E_{int}$) energies when the MOF is treated as rigid in MOF+adsorbate relaxations. We define $E_{int}$ separately from $E_{static}$ because $E_{int}$ is used to conveniently decompose $E_{ads}$ into MOF-adsorbate interactions and MOF deformation contributions, but $E_{static}$ is the quantity used as the adsorption energy in the majority of the current literature on gas separations in MOFs.

\subsection{Energy calculations}
All adsorption energy calculations were performed in a similar manner as in ODAC23. Empty MOFs were first relaxed with the unit cell parameters included as degrees of freedom. Adsorbate molecules were then placed according to the ODAC23 placements, and the combined geometry was relaxed while holding cell parameters fixed to give $E_{sys}(r_{sys})$. The adsorbed-state MOF energy, $E_{MOF}(r_{sys})$, was then determined by removing the adsorbate from the combined system and computing the single point energy of the MOF. In a departure from the method used on ODAC23, we then relaxed the empty MOF starting from its adsorbed-state geometry using fixed unit cell dimensions. The resulting energy was compared to that from the original empty MOF relaxation, and the lower (more favorable) energy structure was taken to be the ground state reference MOF with energy $E_{MOF}(r_{MOF})$. Adsorbates were placed in the ground state MOF and relaxed while holding the MOF atom positions fixed for $E_{static}$ calculations. The gas-phase adsorbate reference energies, $E_{adsorbate}(r_{adsorbate})$, for \ce{CO2} and \ce{H2O} were determined from separate DFT relaxations.

Re-relaxing empty MOFs following  MOF\discretionary{+}{}{+}adsorbate relaxation is a necessary step to avoid overestimating binding energies. This additional relaxation step is necessary when the adsorbate molecule breaks symmetry and allows the MOF to relax more fully compared to the relaxation of the empty MOF alone. In these examples, the new lower energy structure is a better representation of the true structure of the MOF than the structure taken from the CoRE MOF dataset. The ODAC23 data set included hundreds of examples where re-relaxation of this kind was found to lead to lower energy empty MOF structures than the original DFT-relaxed structure.\cite{ODAC23} This re-relaxation approach was not used systematically, however, in the ODAC23 study for all structures. One possible origin of this outcome is that the original crystal structures were solved using simplifying symmetries that are not actually present in the lower energy structure. In these cases, the physically relevant adsorption energy would require using the lower-energy empty MOF system as a reference. This step was only performed in ODAC23 for a subset of materials due to the size of the dataset, but the smaller scale of this work allows us to be rigorous for all 60 systems.

Single-point energies for the adsorbate molecules, $E_{adsorbate}(r_{sys,vacuum})$, were calculated in a 20$\times$20$\times$20 Å box to ensure that interactions with periodic images were negligible. Directly using MOF unit cells for adsorbate energy calculations can be problematic because MOF unit cells can be narrow in one direction. Self-interaction across periodic images ($E_{adsorbate,self}$) can be non-negligible in such cases. Fortunately, the adsorbate self-interaction correction energies in this work were small because all of the MOF unit cells included in this study are sufficiently large. Thus, the $E_{adsorbate,self}$ term in Eqn. \ref{eq:Ediff} was treated as negligible relative to the adsorption energies. It is common in HTS calculations to treat small adsorbate molecules as rigid, so we exclude $E_{adsorbate,deform}$ from Eqn. \ref{eq:Ediff} and from all analysis. These assumptions are validated below. Since we seek to compare FFs and because we initialized each geometry using the same coordinates for each FF, neglecting these effects should have no effect on our qualitative conclusions.

Because any FF is not entirely consistent with DFT, computing FF adsorption energies of DFT-determined geometries that do not represent energy minima for the FFs is not practically relevant. To achieve a stringent test of the FFs and to emulate the approach commonly taken in existing literature, our FF energy calculations are all initialized using unrelaxed MOF geometries and adsorbate placements. Every FF energy presented in this work is determined from relaxations using the specified FF.

\subsection{DFT parameters}
DFT simulation parameters and initial adsorbate positions were taken directly from ODAC23, and full calculation details are available in that publication.\cite{ODAC23} Briefly, we used the PBE exchange-correlation functional\cite{PBE} with a D3 dispersion correction,\cite{Grimme_DFTD} a plane wave energy cutoff of 600 eV, and a 1 $\times$ 1 $\times$ 1 k-point grid. All calculations were spin-polarized and were performed in the Vienna Ab Initio Simulation Package (VASP) v5.\cite{VASP}

We initialized empty MOF geometries here using the ODAC23 relaxed empty MOF geometries. Thus, most empty MOF relaxations converged quickly. A handful of MOF relaxations, however, did not converge quickly due to significant changes in the unit cell parameters relative to ODAC23. This occurred when empty MOFs were re-relaxed in the ODA23 work because the presence of a guest molecule in one of the ODAC MOF+adsorbate relaxations broke the initial MOF structure's symmetry and yielded a lower-energy ground state empty MOF. The re-relaxations in ODAC23 for these examples did not relax unit cell parameters as we do here, so some differences arise between this work and the ODAC23 work for such MOFs.

\subsection{Classical force field parameters}
Our classical force field approach combined UFF4MOF\cite{UFF4MOF1,UFF4MOF2} for all MOF degrees of freedom with point charges from the density-derived electrostatic charges (DDEC) method\cite{DDEC} taken directly from the ODAC23 work. Point charges on MOF atoms were only used in computing MOF-adsorbate interactions since UFF4MOF does not include charge contributions. TraPPE\cite{Potoff_CO2, TraPPE} and the SPC/E model\cite{SPCE} were used for \ce{CO2} and \ce{H2O}, respectively. The SPC/E model was chosen over more complex water models such as TIP4P\cite{TIP4P} because of its simplicity. The ability to model water without placing massless sites was useful because the adsorbate configurations were taken directly from the ODAC23 work, which does not include massless sites.\cite{Chen2000,Zielkiewicz2005} Lorentz-Berthelot mixing rules\cite{Lorentz1881,Berthelot1889} were used to define Lennard-Jones interactions. A cutoff of 12.5 Å without tail corrections was used for all UFF calculations, and long-range electrostatics were treated using an Ewald summation with a force accuracy of $10^{-5}$
kcal/mol/Å.\cite{Ewald1921} While numerous other classical FFs have been used in HTSs to define MOF-adsorbate interactions,\cite{Boyd2017,Cleeton2023} UFF4MOF is one of the few general-purpose classical FFs available to describe MOF degrees of freedom in flexible materials.\cite{UFF4MOF1,UFF4MOF2,ZhenziYu2021} The classical methods we selected here are expected to be largely representative of existing literature. 

All classical FF calculations were performed in the Large-Scale Atomic/Molecular Massively Parallel Simulator (LAMMPS) package,\cite{LAMMPS} with LAMMPS input files generated using \texttt{lammps-interface}.\cite{Boyd2017} All LAMMPS input and log files are available in the SI. Relaxations were performed using the Polak-Ribiere conjugate gradient algorithm\cite{PolakRibiere} with a force cutoff of 1.153 kcal/mol/Å (0.05 eV/Å) to match that from the DFT calculations. The maximum of the atomic forces was used to determine the force convergence rather than the default 2-norm of the global force vector, and a quadratic line search algorithm was used. For the empty MOFs, we performed 200 iterations of atomic coordinate relaxation followed by triclinic cell relaxation to avoid converging to local minima. The maximum volume change per iteration was set to 10\% to aid convergence for one empty MOF relaxation exhibiting pressure instability (ID 9\_66). MOF+adsorbate relaxations consisted of 10 iterations of geometry relaxation in a fixed unit cell. A maximum of 5,000 energy evaluations and 50,000 force evaluations per iteration were allowed for all simulations. 

\subsection{ML force field parameters}
All ML force fields were used with default parameters. The M3GNet model was imported using the Materials Graph Library (MatGL) package with a default radial cutoff was 5 Å.\cite{M3GNet} CHGNet also used a cutoff of 5 Å, and the MACE models and eSEN all used a radial cutoff of 6 Å. The EqV2-ODAC model used a radial cutoff of 8 Å and a maximum of 20 neighbors per atom.\cite{Equiformer_arch,ODAC23} Unit cell relaxations for empty MOFs using MLFFs were performed using Python's Atomic Simulation Environment (ASE)\cite{ASE} \texttt{FrechetCellFilter}. The empty MOF unit cell was not relaxed with EQV2-ODAC because the model does not predict stresses; the unit cell was also fixed for all MOF+adsorbate relaxations. 

All MLFF relaxations were performed in ASE using the Broyden-Fletcher-Goldfarb-Shanno (BFGS) algorithm with a force tolerance of 0.05 eV/Å, a maximum atomic displacement per iteration of 0.05 Å, and a maximum of 1,000 iterations. As with DFT and UFF, MLFF relaxation energies for gas-phase \ce{CO2} and \ce{H2O} were determined via geometry optimization in a 20$\times$20$\times$20 Å box with each respective MLFF. Although no MLFFs were trained with data from small molecules, our isolated molecule relaxations produced reasonable \ce{CO2} and \ce{H2O} structures that were consistent with those from DFT. All Python scripts and log files are available in the SI.

\subsection{Initial geometries for FF relaxations}
Since we are interested in the ability of FFs to self-consistently compute adsorption energies, we used unrelaxed CoRE MOF\cite{CoRE_MOF,CoRE2019} structures with coordinates $r_{MOF}^{CoRE}$ to initialize most empty MOF geometries. The CoRE superscript represents unrelaxed MOF coordinates taken directly from the CoRE MOF database.\cite{CoRE_MOF, CoRE2019} This was the same approach used in the ODAC23 work. In systems involving re-relaxed MOFs as described above for DFT calculations, we initialized MOF positions with the same positions used to initialize the ODAC23 re-relaxations. This ensures a fair comparison between our DFT and FF calculations; the systems affected by re-relaxation are noted by an asterisk in Table \ref{tab:systems}.

In all $E_{ads}$, $E_{int}$, and $E_{static}$ calculations, initial adsorbate molecule placements were taken from ODAC23. Since relaxed unit cell parameters can differ between the FFs, the relative adsorbate molecule position was maintained to ensure a fair comparison between different models despite different FF-determined geometries for a given MOF. Specifically, the fractional coordinates of the molecule’s center of mass were retained, and the bond lengths and molecular orientation were identical to those in ODAC23. Adsorbate molecule positions were not adjusted for DFT simulations where the relaxed unit cell dimensions in this work closely matched those from ODAC23.
\section{Results and Discussion}

\subsection{Dataset selection} 
The ODAC23 dataset includes DFT data for more than 8,000 MOFs, including more than 3,400 structures containing point defects. Thoroughly implementing and testing UFF4MOF for a diverse range of structures is time-consuming when not assuming rigidity because LAMMPS structural file generation is nontrivial for large systems such as MOFs. Thus, we selected a relatively small subset of the ODAC23 materials for our calculations. This also reduced the overall computational cost.

\begin{figure*}[ht!]
    \centering
    \includegraphics[width=\textwidth]{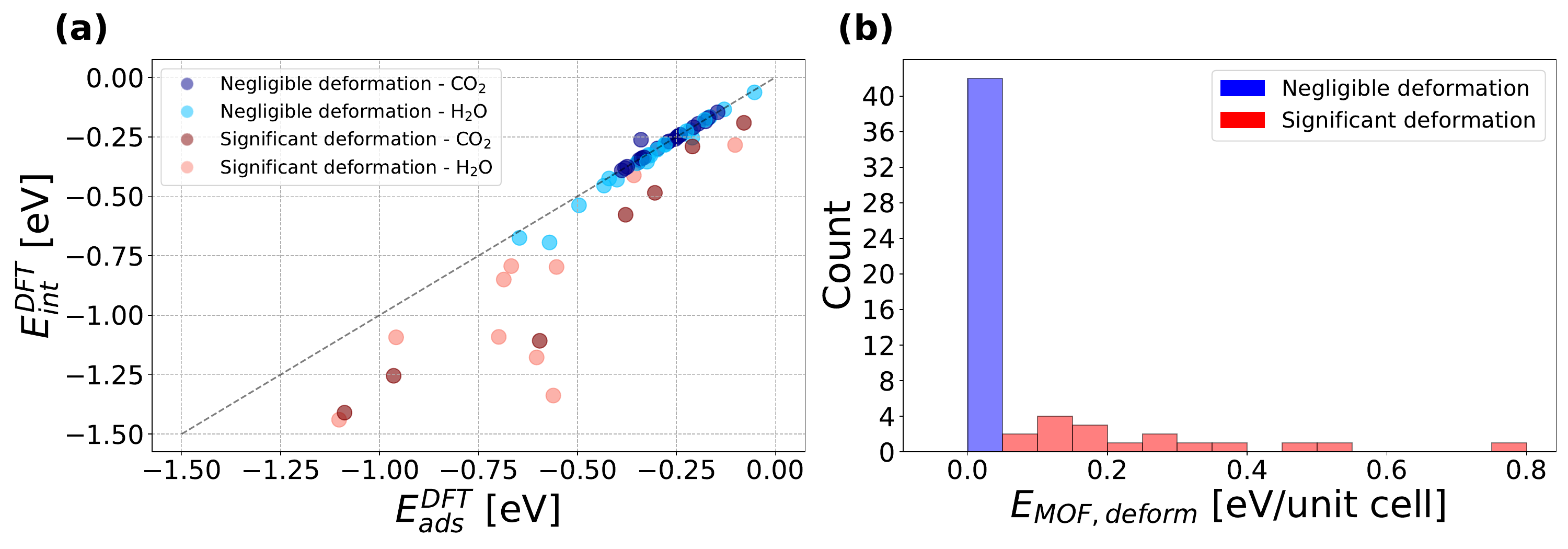}
    \caption{(a) Distribution of DFT adsorption and interaction energies in the 59 selected MOF + adsorbate systems. (b) Histogram of MOF deformation energies colored by deformation class.}
    \label{fig:dataset}
\end{figure*}

A total of 60 MOF+adsorbate systems were taken from the ODAC23 in-domain ``initial structure to relaxed energy'' (IS2RE) test set for further study in this work. Only the test set was considered to avoid data leakage in the EqV2-ODAC predictions, and only systems with a single adsorbate molecule, either \ce{CO2} or \ce{H2O}, were considered for simplicity. First, the MOFs in the IS2RE test set were screened for their DAC favorability according to ODAC23 DFT calculations. A MOF is considered favorable for DAC if, for all ODAC23 IS2RE single-adsorbate train, validation, and test MOF+adsorbate systems, the system with the most favorable DFT adsorption energy contains \ce{CO2} and not \ce{H2O}. This is an unusual phenomenon within the complete spectrum of MOFs that exist, but it is highly desirable for DAC.\cite{Findley2021} Similarly, MOFs are also considered potentially favorable for DAC if the average adsorption energy for systems containing \ce{CO2} is more favorable (i.e., more negative) than that for systems containing \ce{H2O}.

We used an algorithm to identify systems for this study, but some manual curation on a handful of systems was required to ensure a diverse and balanced dataset of interesting materials. The algorithm used to select systems can be found in the SI. MOFs with an absolute DFT-calculated $E_{MOF,deform}$ $\leq$ 0.05 eV per unit cell were considered to undergo negligible deformation, while MOFs with $E_{MOF,deform}$ $>$ 0.05 eV per unit cell were considered to undergo significant deformation. Systems were added to the data set to approximately balance the two deformation classes and adsorbate types while prioritizing DAC favorability and avoiding duplicates of MOFs.

One system (ID $3\_78$) was excluded from further study due to unit cell contraction and pore collapse upon DFT relaxation  (see Figure \ref{fig:pore_collapse}), leaving a total of 59 systems in the dataset. This material was re-relaxed with fixed unit cell parameters in ODAC23. When including cell parameters as degrees of freedom in this work, the resulting MOF geometry significantly differs from the geometry included in ODAC23. We calculated a positive DFT adsorption energy for \ce{CO2} in this material, yielding it unsuitable for adsorptive separations.

The 59 systems are split between \ce{CO2} and \ce{H2O} and between the negligible and significant deformation classes. Our DFT calculations include some differences from ODAC23, so the classes are not balanced because dataset selection was based on ODAC23 DFT energies. Re-relaxation of empty MOFs after removing the guest molecule led to multiple examples where MOF deformation would have been classified as significant using ODAC23 data but were reclassified as negligible MOF deformation cases once lower energy empty MOF structures were found in our current calculations. Of the 42 systems with negligible deformation, 22 (20) contain \ce{CO2} (\ce{H2O}), and of the 17 systems with significant deformation, 7 (10) contain \ce{CO2} (\ce{H2O}). 

The ODAC23 dataset includes information for large numbers of pristine MOFs and MOFs containing linker vacancies. Our dataset here contains 31 systems derived from 28 pristine MOFs and 28 systems derived from 23 defective MOFs. About two-thirds of the systems contain MOFs that are potentially favorable for DAC according to our above definitions using the ODAC23 adsorption energies, with 20 very favorable systems, 19 potentially favorable systems, and 20 not favorable systems according to ODAC23 calculations. The systems not favorable for DAC were included because we could not identify 60 systems from the test set that satisfied the DAC favorability criteria. A summary of MOF+adsorbate systems we considered is given in Table \ref{tab:systems}. Figure \ref{fig:dataset}(a) shows an overview of the dataset split into negligible deformation (blue) and significant deformation (red) classes. Markers are colored with darker and lighter shades to denote the presence of a \ce{CO2} or \ce{H2O} guest molecule, respectively. According to Equation \ref{eq:Ediff}, larger deviations from the parity line indicate larger MOF deformation energies. Our dataset also represents a mix of physisorption and chemisorption with 44 and 15 systems, respectively, where chemisorption is defined as a DFT $E_{int}$ $<$ –0.5 eV. We use $E_{int}$ to define chemisorption since it directly probes the contribution of electronic interactions that underlie the covalent bonding characteristic of chemisorption. $E_{ads}$ remains the most relevant quantity in this work, but MOF deformation can weaken $E_{ads}$ even when chemisorption is occurring as indicated by $E_{int}$.

Figure \ref{fig:dataset}(b) shows the corresponding MOF deformation energies. Positive deformation energies indicate that relaxation in the presence of the adsorbate molecule yields a MOF structure that is less energetically favorable than the empty MOF relaxed alone (as in Figure \ref{fig:deform_vis}). Upon removal of the adsorbate molecule, these MOFs return to their empty ground state. The one point noticeably above the parity line in Fig. \ref{fig:dataset}(a) is system 1\_230, which has a $E_{MOF,deform}$ of 0.002 eV but lies above the parity line due to its attractive $E_{adsorbate,self}$ of –0.08 eV.

Concerns regarding the validity of MOFs from computation-ready databases have recently arisen, and prior studies have suggested partial charge assignments can used as a surrogate for bond orders to screen MOFs for misbonded atoms.\cite{ChenManz2020,White2024,Gibaldi2025,CoRE2025} We used the MOFChecker software\cite{MOFChecker} to confirm partial charge assignments from the EQeq method\cite{Wilmer2012} are physically reasonable in every system we examined. For additional future study, DDEC charges derived from the ODAC23 DFT calculations are also available in that database. DDEC charges are expected to be more accurate than the charge equilibration methods used in other publications,\cite{DDEC} and examples are known where DDEC charges derived from DFT calculations are physically reasonable even when EQeq charges are not.\cite{Nazarian2016}
\subsection{Energy calculations}

We first validate our assumptions regarding adsorbate self-interaction ($E_{adsorbate,self}$) and deformation ($E_{adsorbate,deform}$). In our calculations only two DFT adsorbate self-interaction energies exceeded 0.01 eV with a maximum of 0.079 eV. Thus, we neglect $E_{adsorbate,self}$ from Eqn. \ref{eq:Ediff} in our analysis. Likewise, the maximum and average DFT $E_{adsorbate,deform}$ in our dataset are 0.078 eV and 0.0025 eV, respectively. The largest adsorbate deformation energies represent only minor changes in the geometry of the adsorbate molecule (e.g., O-C-O angle). In the case with the largest adsorbate deformation energy of 0.078 eV, the corresponding MOF deformation energy is 0.50 eV, meaning that only 13\% of the total deformation energy is due to the adsorbate. In total, 50 of 59 systems have an absolute DFT adsorbate deformation energy of less than 0.01 eV. For these reasons, we lump $E_{adsorbate,deform}$ in Eqn. \ref{eq:Ediff} into the $E_{MOF,deform}$ term with the understanding that MOFs contribute almost all of the deformation energy for a given system. 

\subsubsection{Adsorption in rigid MOFs}
\begin{figure*}[ht!]
    \centering
    \includegraphics[width=\textwidth]{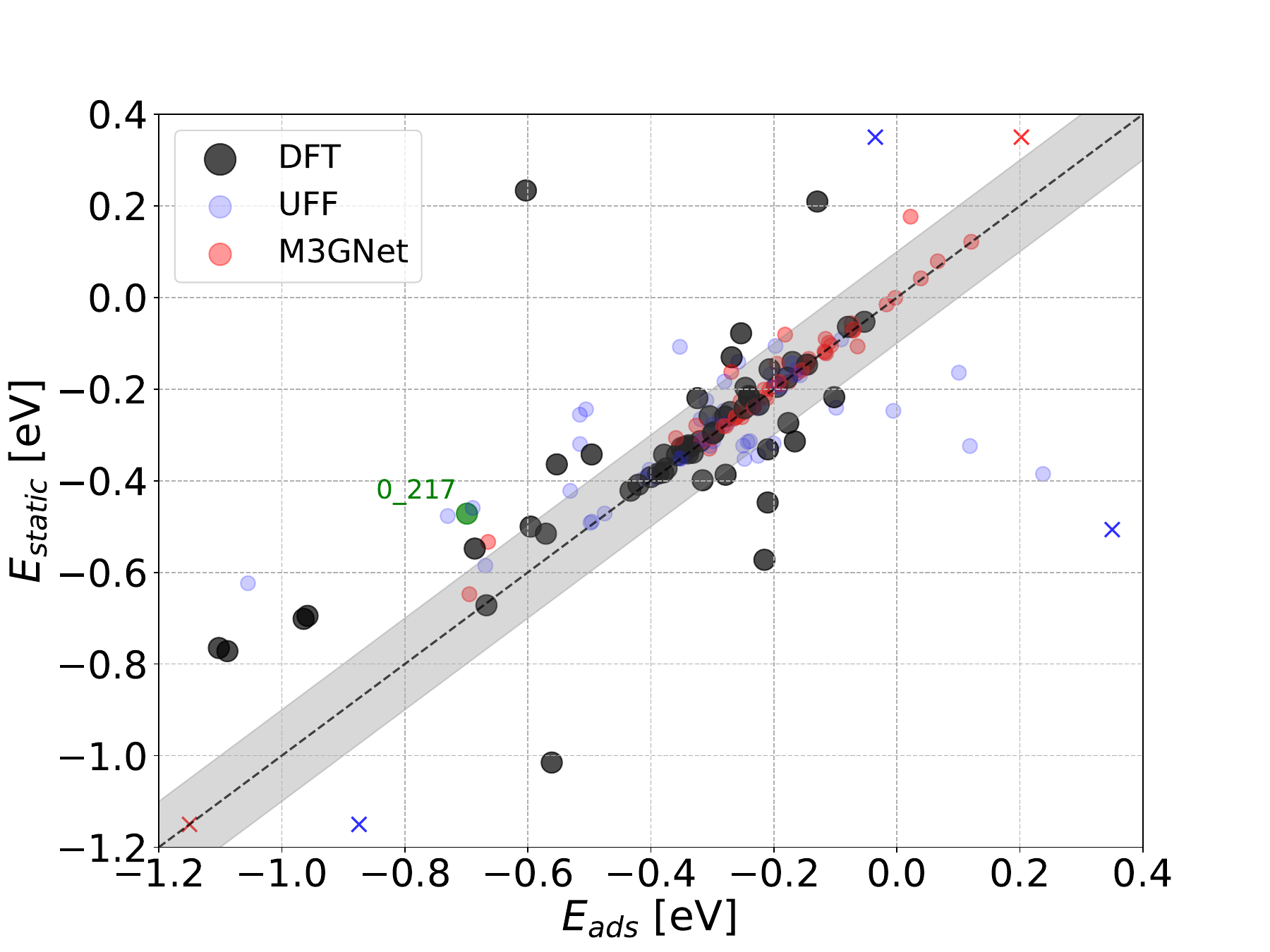}
    \caption{Comparison of $E_{ads}$ with fully flexible MOFs to $E_{static}$ in rigid MOFs using DFT, UFF, and M3GNet. System 0\_217 from Figure \ref{fig:deform_vis} is highlighted in green. X markers indicate a point lying beyond the bounds of the plot. The shaded area within $\pm$0.1 eV indicates the region in which good agreement between $E_{ads}$ and $E_{static}$ is achieved.}
    \label{fig:GCMC_compare}
\end{figure*}

We begin our analysis by considering whether MOF flexibility makes a meaningful difference in adsorption energies for simulations commonly performed in the literature, namely GCMC simulations with rigid MOFs. Figure \ref{fig:GCMC_compare} compares adsorption energy calculations in flexible MOFs to $E_{static}$ using DFT, UFF, and M3GNet. Each $E_{ads}$ and $E_{static}$ is computed using the mode specified by the point coloring. The DFT points in Figure \ref{fig:GCMC_compare} are shown in bold to emphasize that DFT is taken as the ground truth in this study. The result for system 0\_217, the example shown in Figure \ref{fig:deform_vis}, is highlighted in green.

Systems that do not significantly deform naturally fall close to the parity line in Figure \ref{fig:GCMC_compare}. Many points fall further from the parity line for all three models, particular in the chemisorption regime from $-1.2 $ eV$ < E_{int} < -0.5$ eV where most chemistries promising for DAC occur. The deviation between $E_{static}$ and $E_{ads}$ is considerable at the DFT level, indicating these deviations are not simply shortcomings of the FFs. Including all 59 systems, the MAE between these two quantities computed with DFT is 0.268 eV when -1.2 $<$ $E_{int}$ $<$ -0.5 eV and 0.056 eV when -0.5 $\le$ $E_{int}$ $<$ 0.0 eV. These results indicate that there are many examples in which neglecting adsorbate-induced MOF flexibility leads to significant errors in determining the adsorption affinity of molecules.

The scatter in the DFT points in Figure \ref{fig:GCMC_compare} is the key takeaway of this section, and the UFF and M3GNet results are provided only for comparison. Figure \ref{fig:rigid_scatter_vs_DFT} shows UFF and M3GNet $E_{static}$ compared to DFT $E_{ads}$. While M3GNet appears less susceptible to the effects of MOF flexibility in Figure \ref{fig:GCMC_compare}, the scatter in Figure \ref{fig:rigid_scatter_vs_DFT} highlights M3GNet's poor performance relative to the ground truth DFT results. That is, our results imply that using M3GNet under a rigid MOF assumption is often a poor approximation for fully flexible DFT $E_{ads}$ calculations. 

Figure \ref{fig:GCMC_by_system} shows that the MOFs for which the rigidity assumption is invalid according to one FF differ from those identified using other FFs. There is also no correlation between the suitability of the rigidity assumptions and the magnitude of MOF deformation and pore limiting diameter (PLD) of the relaxed empty MOF. That is, it is not straightforward to know \textit{a priori} for which MOFs the rigidity assumption is reasonable. These results hint that many of the materials of greatest interest for DAC cannot readily be modeled using rigid MOF structures. This observation motivated us to examine the underlying causes of the systematic errors among the different FFs in more depth.

\subsubsection{Classical force field (UFF)}

\begin{table*}[ht!]
    \centering
    \renewcommand{\arraystretch}{1.0}
    \setlength{\tabcolsep}{5pt}
    \resizebox{0.85\linewidth}{!}{
    \tiny
    \begin{tabular}{lccccc}          
        \toprule
        Force field & \ Subset & \ $E_{ads}$ & \ $E_{static}$ & \ $E_{int}$ & \ $E_{MOF,deform}$ \\
        \midrule
                \multirow{3}{*}{UFF} & Physisorption & 0.156 & 0.085 & 0.168 & 0.025 \\
                             & Chemisorption & 0.380 & 0.462 & 0.682 & 0.358\\
                             & Total & 0.213 & 0.181 & 0.299 & 0.109\\ 
        \midrule
                \multirow{3}{*}{M3GNet} & Physisorption & 0.136 & 0.164 & 0.128 & 0.020 \\
                             & Chemisorption & 0.575 & 0.541 & 0.799 & 0.205 \\
                             & Total & 0.247 & 0.260 & 0.298 & 0.067 \\ 
        \midrule
                \multirow{3}{*}{CHGNet} & Physisorption & 0.092 & 0.160 & 0.095 & 0.018 \\
                             & Chemisorption & 0.217 & 0.271 & 0.362 & 0.202 \\
                             & Total & 0.124 & 0.189 & 0.163 & 0.065 \\ 
        \midrule
                \multirow{3}{*}{MACE-MP-0} & Physisorption & 0.161 & 0.224 & 0.121 & 0.057 \\
                             & Chemisorption & 0.238 & 0.394 & 0.364 & 0.210 \\
                             & Total & 0.181 & 0.268 & 0.183 & 0.096 \\ 
        \midrule
                \multirow{3}{*}{MACE-MPA-0} & Physisorption & 0.365 & 0.304 & 0.366 & 0.023 \\
                             & Chemisorption & 0.334 & 0.428 & 0.288 & 0.149\\
                             & Total & 0.357 & 0.336 & 0.346 & 0.055\\ 
        \midrule
                \multirow{3}{*}{eSEN} & Physisorption & 0.187 & 0.224 & 0.192 & 0.019 \\
                             & Chemisorption & 0.409 & 0.449 & 0.637 & 0.221\\
                             & Total & 0.243 & 0.281 & 0.305 & 0.070\\ 
        \midrule
                \multirow{3}{*}{EqV2-ODAC} & Physisorption & 0.182 & -- & -- & -- \\
                             & Chemisorption & 0.220 &  -- & -- & -- \\
                             & Total & 0.191 & -- & -- & -- \\ 
        \bottomrule
    \end{tabular}
    }
    \caption{Mean absolute errors in eV relative to DFT split between physisorption and chemisorption for each energy term using each FF in this study. There are 44 and 15 systems in the physisorption and chemisorption sets, respectively.}
    \label{tab:MAE_tab}
    \vspace{-10pt}
\end{table*}

\begin{figure*}[ht!]
    \centering
    \includegraphics[width=\textwidth]{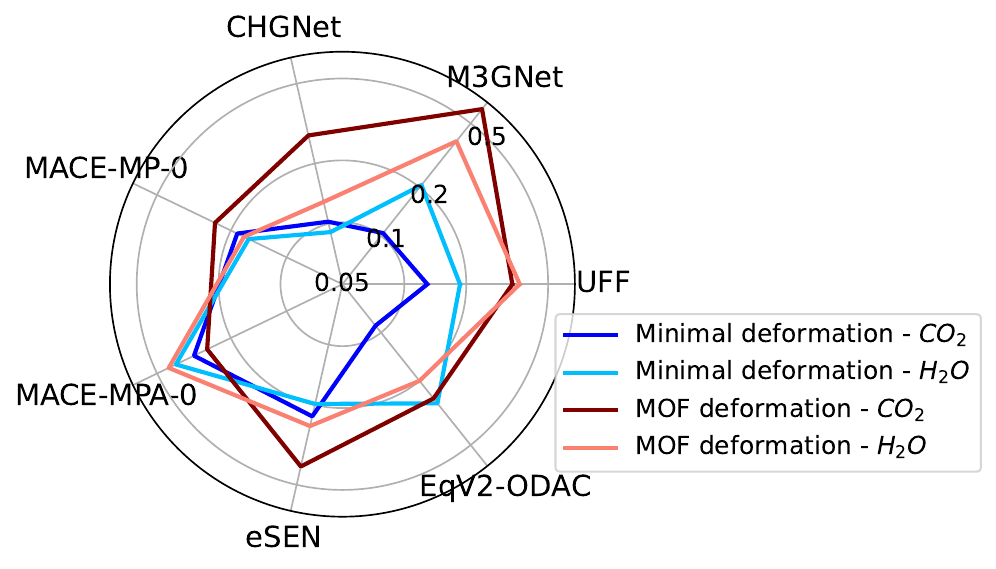}
    \caption{Mean absolute adsorption energy errors relative to DFT in eV separated by deformation class for FF relaxations. The radial axes show energies on a logarithmic scale, with the origin representing a MAE of 0.05 eV.}
    \label{fig:radar}
\end{figure*}

We next consider the performance of our classical FF for describing adsorption when MOFs are free to deform. The resulting mean absolute errors (MAEs) in adsorption energies for all 59 systems are shown in Table \ref{tab:MAE_tab} and Figure \ref{fig:radar}. The DFT energies are taken as the ground truth for error calculations. UFF is better at describing MOFs with negligible deformation energies then MOFs with significant deformation energies. Table \ref{tab:MAE_tab} shows that UFF struggles to describe chemisorption, defined here as DFT $E_{int}$ $<$ –0.5 eV. This is unsurprising because UFF is not a reactive FF. We will see below, however, that UFF approaches the adsorption energy MAE of the best performing MLFFs when only considering physisorption, and UFF outperforms the MLFFs when the MOF is assumed rigid ($E_{static}$ in Table \ref{tab:MAE_tab}). That is, UFF may be a suitable choice for describing phyisorption in rigid MOFs, the context in which it is most often used. Figure \ref{fig:GCMC_compare}, however, reminds us that the rigidity assumption is frequently invalid and that UFF cannot be used in a simple way to detect this outcome.

We reiterate that every FF adsorption energy calculation was performed self-consistently, so the empty MOFs used in the FF relaxations differ from those used in DFT. This stringent test of the FFs is intended to mimic a standard FF adsorption energy workflow without bias from DFT simulations. A comparison of UFF adsorption energy predictions to DFT in only DFT-relaxed MOFs is shown in the ODAC23 work, and we found very good agreement between the two methods in the physisorption regime.\cite{ODAC23}

\begin{figure*}[ht!]
    \centering
    \includegraphics[width=\textwidth]{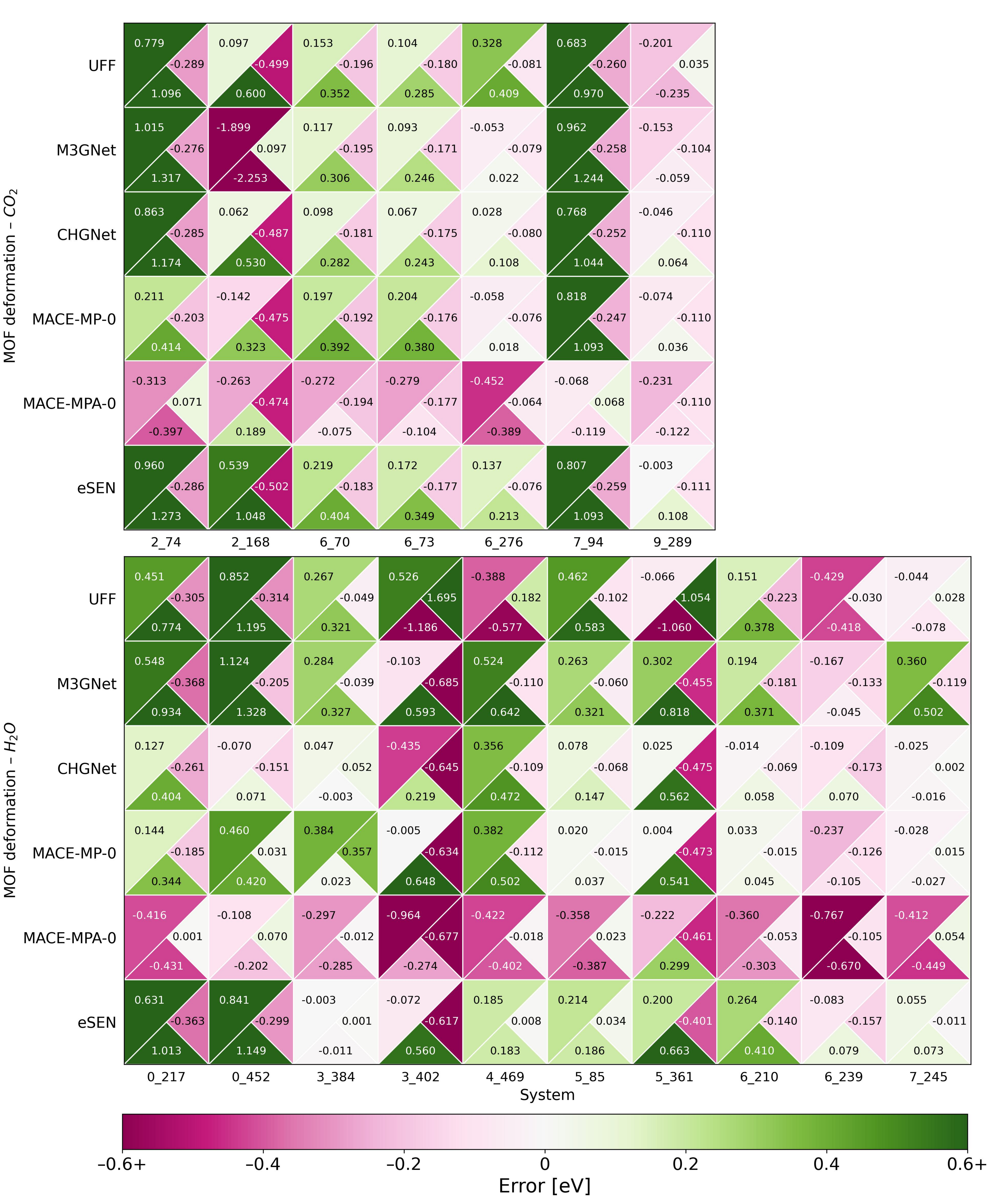}
    \caption{Adsorption energy errors (upper left triangle), interaction energy errors (bottom triangle), and MOF deformation errors (right triangle) from FF relaxations for systems that undergo significant deformation separated by adsorbate. All energies are in eV using the color bar shown in the figure.}
    \label{fig:gridplot}
\end{figure*}

The decomposition of adsorption energy into interaction and deformation energies in Eqn. \ref{eq:Ediff} presents a useful way to further explore the cause of adsorption energy errors. A low adsorption energy error may be the result of cancellation of errors between $E_{int}$ and $E_{MOF,deform}$, so it is not enough to only consider adsorption energy when evaluating the accuracy of FFs. FFs suitable for predicting adsorption behavior in non-rigid MOFs must adequately describe both MOF–adsorbate interactions and MOF deformation. Figure \ref{fig:gridplot} shows signed adsorption energy errors labeled by system ID and separated by adsorbate with the corresponding interaction and MOF deformation energy errors. Only systems undergoing significant deformation are shown in Fig. \ref{fig:gridplot}; the same plot for systems that undergo negligible deformation is shown in Figure \ref{fig:gridplot_supp}. The coloring of Figs. \ref{fig:gridplot} and \ref{fig:gridplot_supp} shows larger errors relative to DFT as darker shades of pink (negative) and green (positive). Cancellation of errors occurs when the right triangle ($E_{MOF,deform}$) and the bottom triangle ($E_{int}$) differ in color, signifying a positive error relative to DFT for one quantity and a negative error for the other. The resulting adsorption energy error is then lower than the maximum of the interaction and MOF deformation energy errors. An example of this is system $2\_74$ for all five FFs. 

UFF errors are shown on the first row in Figures \ref{fig:gridplot} and \ref{fig:gridplot_supp}. Interaction energy errors are consistently large, indicating that UFF fails to adequately describe non-bonded interactions and, in the case of chemisorption, chemical bonding between the host and guest molecules. UFF also performs poorly when predicting MOF deformation energies for MOFs that significantly deform according to DFT, as indicated by the darker shades of color in the right triangles. Overall, there is little discernible pattern across the UFF errors, which is consistent with the idea that the classical FF is sensitive to slight changes in atomic positions and electrostatics. The UFF adsorption energy MAE of 0.213 eV is almost twice that of the best performing MLFF.

The top row of Figs. \ref{fig:gridplot} and \ref{fig:gridplot_supp} also shows that cancellation of errors is common in our UFF calculations, occurring in 42 of 59 systems. Significant cancellation of errors in nearly half of systems, meaning that UFF often performs more poorly describing interaction energy or MOF deformation energy than it does describing the overall adsorption energy.

\subsubsection{Foundational MLFFs}

The MAEs for the four foundational MLFFs -- M3GNet, CHGNet, both MACE models, and eSEN -- are shown in Table \ref{tab:MAE_tab} and Figure \ref{fig:radar}. The $E_{ads}$ MAEs for M3GNet, CHGNet, and MACE-MP-0 outperform that of UFF, but MACE-MPA-0 severely underperforms and exhibits the largest MAE of any FF tested. CHGNet and MACE-MP-0 perform best of all total energy models tested with moderate to significant improvements in $E_{int}$ and $E_{MOF,deform}$ accuracy relative to both UFF and M3GNet. Like all the MLFFs, CHGNet and MACE-MP-0 struggle with chemisorption compared to physisorption, but the performance disparity between the two adsorption types are lower than for other FFs.

It is not surprising some MLFFs outperform UFF because they were trained on databases of DFT calculations. Even still, percent errors in adsorption energy approach and, in some cases, exceed 100\%. To determine whether these errors are reasonable, consider the published test set errors for CHGNet and for MACE-MP-0. CHGNet achieves an MAE of 33 meV/atom on its MPtrj test set,\cite{Deng2023} and MACE-MP-0 achieves an MAE of 33 meV/atom for MOF total energy predictions.\cite{MACE-MP-0} The MOFs in our dataset range from 66 to 492 atoms/unit cell with an average of 187, corresponding to average total energy errors from 2.2 to 16 eV per unit cell with both CHGNet and MACE-MP-0. While much of these errors may cancel during an adsorption energy calculation, it is not unreasonable to find resulting $E_{ads}$ errors on the same magnitude of $E_{ads}$ itself. That is, the good performance of MLFFs on their test sets does not necessarily translate to low percent errors in adsorption energy.

The relatively poor performances of MACE-MPA-0 and eSEN given their sizes are unexpected and should be further investigated, but why MLFFs perform well or poorly is beyond the scope of the current work. Both FFs contain more parameters and were trained on more structures than the other models considered in this work. We hypothesize that the inclusion of sAlex (+ OMat for eSEN) in the training data and the size of the FFs make the models overfitted and unsuited for describing adsorption in MOFs as performed in this work.


The decomposition of errors is shown in rows 2–6 of each block in Figures \ref{fig:gridplot} and \ref{fig:gridplot_supp}. Cancellation of errors occurs in 33 systems for M3GNet, 37 systems for CHGNet, 28 systems for MACE-MP-0, 39 systems for MACE-MPA, and 33 systems for eSEN. The ability of foundational MLFFs to describe van der Waals interactions with reasonable accuracy is particularly interesting and is worthy of further study given the lack of training data that includes dispersion effects in the MP database. Despite improvement over UFF, $E_{ads}$ errors remain significantly greater than the 0.1 eV target discussed above for all MLFFs, meaning no FF is yet accurate enough for reliably studying adsorption in MOFs.

\subsubsection{ODAC23 Equiformer (EqV2-ODAC)}

Table \ref{tab:MAE_tab} and Figure \ref{fig:radar} show that EqV2-ODAC performed similarly to both CHGNet and MACE-MP-0. EqV2-ODAC is excluded from some columns of Table \ref{tab:MAE_tab} and from Figures \ref{fig:gridplot} and \ref{fig:gridplot_supp} because the model directly predicts adsorption energy and therefore cannot decompose errors into interaction and MOF deformation contributions. We conclude that EqV2-ODAC successfully learned the error cancellation implicitly from DFT. It is also interesting that the top-performing MLFFs rival EqV2-ODAC given the latter's underlying training data exclusively containing \ce{CO2} and \ce{H2O} adsorption in MOFs and its more than 153 million parameters.

\subsubsection{Deformation as a classification problem}

\begin{figure*}[ht!]
    \centering
    \includegraphics[width=\textwidth]{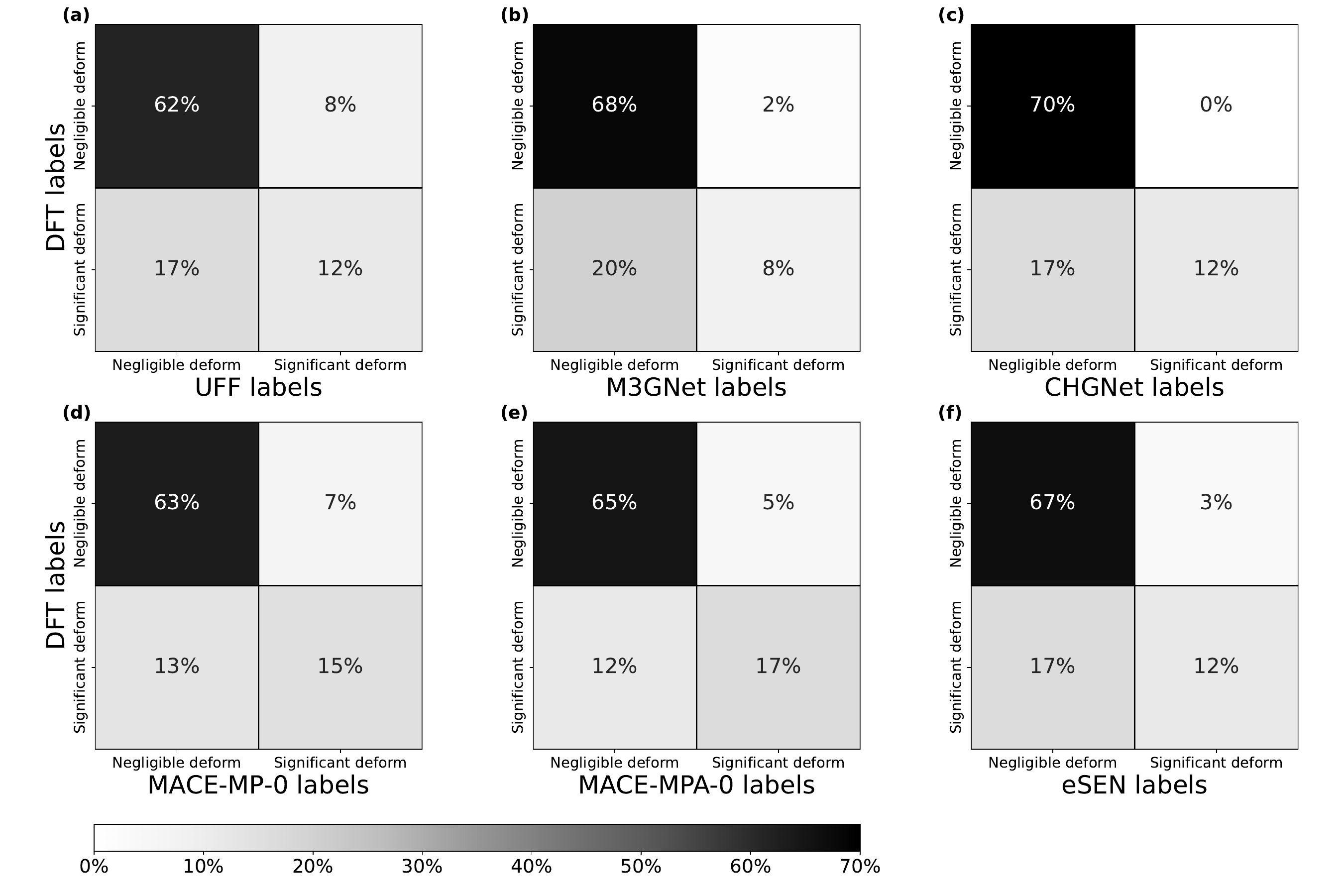}
    \caption{Confusion matrices for deformation class identification for (a) UFF, (b) M3GNet, (c) CHGNet, (d) MACE-MP-0, (e) MACE-MPA-0, and (f) eSEN. True positives and true negatives are located in the upper left and lower right of each matrix, respectively.}
    \label{fig:confusion}
\end{figure*}

We have established that existing MLFFs can outperform a classical FF for describing adsorption in deformable MOFs in terms of MAE, although neither approach has sufficient accuracy for detailed calculations. It is also useful to address whether MLFFs can outperform classical FFs in determining whether a MOF will undergo significant deformation upon the introduction of a guest molecule. We used a threshold of 0.05 eV to classify systems as having significant MOF deformation or not. The classification is performed implicitly based on the FF results, and no explicit classification model is used. As above, DFT results are taken as the ground truth for this analysis.

Figure \ref{fig:confusion} shows the confusion matrices for this classification task based on relaxation calculations for each FF. EqV2-ODAC was again omitted because it cannot decompose $E_{ads}$. All six FFs notably performed very similarly, and there is not one clear best model based solely on this analysis. CHGNet and MACE-MPA-0 both misclassified only 17 \% of the dataset despite MACE-MPA-0 performing worst in terms of $E_{ads}$ MAE. This is consistent with most MACE-MPA-0 energy errors resulting from the $E_{int}$ term in Eqn. \ref{eq:Ediff}. When only considering MOFs that undergo significant deformation, all FFs except MACE-MP-0 and MACE-MPA-0 misclassify more than half of the systems as undergoing negligible deformation. That is, the FFs tend to underpredict MOF deformation where DFT indicates the phenomenon is important. These findings indicate that energy MAEs are not necessarily correlated to whether an FF is able to identify which systems undergo significant MOF deformation. Although MLFFs can outperform UFF in terms of energy MAE, more work is needed to tailor FFs that can adequately describe adsorbate-induced MOF deformation and that can readily identify systems in which such deformation is significant.

A natural question regarding these results is whether large errors were associated with specific systems that were challenging to model. Figure \ref{fig:err_by_system} in the SI shows that systems with outlier errors for one model rarely yielded large errors for other models. That is, the errors were not the result of specific geometries across the four FFs tested. Figure \ref{fig:volumes} in the SI shows unit cell volume errors from FF relaxations relative to those from DFT. These results were consistent with the energy findings presented here, with the MLFFs generally describing unit cell geometry changes better than UFF. 
\subsubsection{Limitations of FF energy predictions}

It is of course interesting to ask whether any of the FFs we tested accurately describes the systems we considered. One limitation of the current study is the relatively small sample size compared to the size of typical MOF databases ($\sim$1-10K). Given the relatively small differences in the performance of the FFs tested, the sample size of 60 is likely not sufficient to draw a strong conclusion about which FF is best for MOF deformation. However, the sample size is sufficient to compute meaningful statistics about the overall performance of the models.

Table \ref{tab:R2} reports the $R^{2}$ and Pearson correlation coefficient values for \ce{CO2} and \ce{H2O} adsorption in the flexible MOFs in our dataset. The definition of $R^{2}$ used in this work is

\begin{align}
R^2 = 1 - \frac{\sum_{i} (y_i - \hat{y_i})^2}{\sum_{i} (y_i - \bar{y})^2}
\label{eq:R2}
\end{align}

\noindent where $y$, $\hat{y}$, and $\bar{y}$ are the true target values, FF predicted values, and mean of the true target values, respectively. $R^{2}$ can be negative if the FF predictions are worse than predicting the mean of the DFT $E_{ads}$. Despite promising MAEs for the top performing FFs, only modest correlation exists between the DFT and FF results, even for the best performing FF. Figure \ref{fig:parity} visually confirms this poor correlation for all FFs, especially in the physisorption regime. This poor performance points to complexity of the task, which requires an accurate treatment of both adsorption and chemical bonding within the MOF framework. The magnitude of typical adsorption energies ($\sim$0.4 eV) and deformation energies ($\sim$0.1 eV) are similar to the magnitude of the $E_{ads}$ MAEs of the FFs ($\sim$0.2 eV), making it unsurprising that correlations are often poor. The poor results for $R^{2}$ may be biased by the fact that the examples chosen were selected to over-represent deformation and chemisorption.  However, deformation and chemisorption are particularly relevant for DAC, and the current results suggest that currently available FFs are not robustly capable of describing these phenomena.

A particularly challenging aspect of using FFs to compute adsorption energies is distinguishing between physisorption and chemisorption. The results include several systems where DFT indicates that chemisorption is occurring but where FFs relax to local minima corresponding to physisorption. Table \ref{tab:short_errors} highlights three systems where DFT and FFs predict chemisorption and physisorption, respectively. The distances between the adsorbate and MOF in the DFT-relaxed geometries are less than 2 Å, consistent with the strong DFT adsorption energies. The errors are consistently high due to FFs predicting physisorption, which skews the $R^{2}$ metric. Physisorption energies have smaller magnitudes and variances (0 $\leq$ $E_{ads}$ $<$ 0.5 eV) compared to chemisorption, which has a higher magnitude and variance (0.5 $\leq$ $E_{ads}$ $<$ 1.2 eV). Physisorption is also more prevalent in both ODAC and this work, with 85\% of ODAC23 systems with one guest molecule and with 75\% of the systems in this work having $E_{ads}$ $<$ 0.5 eV. Thus, it is possible for MLFFs to achieve a low MAE by fortuitously guessing physisorption energies. This is supported by the fact that M3GNet, CHGNet, MACE-MP-0, and eSEN all show low MAEs for physisorption despite having no D3 corrections included in their training data. Their strong performance is not necessarily an indication that they are correctly capturing the underlying physics. These findings further illustrate that MAE is not a good metric for assessing the ability of models to predict chemisorption.

Another known challenge associated with with adsorption energy predictions is that they require an accurate reference energy for the gas-phase adsorbate molecule ($E_{adsorbate}$), which can be challenging for MLFFs to compute given their underlying training data. None of the training data for the four general-purpose MLFFs contains gas-phase molecules under a vacuum, though MP does contain \ce{CO2} and \ce{H2O} in a variety of bulk configurations.\cite{Jain2013_MP} In this work, we relied on the FFs to compute $E_{adsorabte}$ despite their known limitations. An alternative approach that improves accuracy is to treat $E_{adsorbate}$ as a free variable when computing the adsorption energies and to regress them to known targets. We determined a correction for the \ce{CO2} and \ce{H2O} gas-phase energies for each FF using linear regression to remove systematic errors between FF and DFT predictions (see Eqn. \ref{eq:gas_corr} in the SI for details). Table \ref{tab:gas_errors} compares MLFF performance using direct MLFF $E_{adsorbate}$ predictions and using this correction scheme. UFF is shown as a reference despite not suffering from the same challenges in principle. While the MAEs decrease, the ranking of the MLFFs remains similar, except for MACE-MPA-0, which becomes the second most accurate FF after correcting $E_{adsorbate}$. This suggests that systematic errors in calculating gas-phase molecular energies can have an impact on the overall accuracy of some models, but they do not affect the conclusions of this work since the rankings of the FFs based on MAE and $R^{2}$ do not significantly change.  
\section{Conclusions}

Deformation and, more generally, framework flexibility are essential for accurately describing and understanding the mechanisms of adsorption in MOFs.\cite{Jawahery2017,Coudert2011,ZhenziYu2021} We have used the Open DAC 2023 dataset to explore the efficacy of a classical FF model and six MLFFs for describing MOF deformation in adsorption of \ce{CO2} and \ce{H2O} for a subset of materials that are potentially favorable for DAC. Our results show that UFF -- the most common FF used in current MOF studies -- is a poor emulator of DFT and does not accurately model chemisorption and MOF deformation in most cases. MLFFs trained on large databases of DFT calculations such as CHGNet and MACE-MP-0 performed better than UFF. Other MLFFs -- namely M3GNet, MACE-MPA-0, and eSEN -- performed similarly to or worse than UFF. More work needs to be done to determine why MACE-MPA-0 performs poorly for describing adsorption in MOFs, and we hypothesize this may be related to the model's underlying data or size. The EqV2-ODAC model trained on the Open DAC 2023 work\cite{ODAC23} also performed well for predicting adsorption energies, which is not surprising given its underlying training data and size. However, these models all performed poorly when treating MOF deformation as a classification problem and still produce mean absolute adsorption energy errors relative to DFT significantly greater than a 0.1 eV target.

We intend for this work to serve as a useful starting point and motivator for future work focused on developing MLFFs for MOF applications. We have performed the calculations here to be similar to what is currently done in the literature so as to accurately represent the way that most MOF modelers use these models in their own work. Adsorbate-induced deformation and framework flexibility must both be adequately described by any ML model seeking to emulate ab initio method performance at a reduced computational cost. Pre-trained MLFFs have improved rapidly for describing a wide range of materials, including MOFs. Despite this recent progress, however, work is still needed to develop MLFFs that can reliably describe the underlying physics of adsorption in solid sorbents relevant for DAC.

\paragraph{Supporting Information} The Supporting information is freely available at URL

\paragraph{Author Contributions} 
LMB and DSS conceived the study, and LMB carried out all calculations. AJM and DSS provided supervision for the study. All co-authors contributed to the manuscript. 

\begin{acknowledgement}
The authors acknowledge Anuroop Sriram (Meta), Dr. Zachary Ulissi (Meta), and Dr. Sihoon Choi (Georgia Tech) for their reivew of this manuscript and for assistance with running the Equiformer V2 model. DSS acknowledges funds from the ORNL LDRD program.

\end{acknowledgement}
\providecommand{\latin}[1]{#1}
\makeatletter
\providecommand{\doi}
  {\begingroup\let\do\@makeother\dospecials
  \catcode`\{=1 \catcode`\}=2 \doi@aux}
\providecommand{\doi@aux}[1]{\endgroup\texttt{#1}}
\makeatother
\providecommand*\mcitethebibliography{\thebibliography}
\csname @ifundefined\endcsname{endmcitethebibliography}  {\let\endmcitethebibliography\endthebibliography}{}

\clearpage
\onecolumn
\section{Supplementary Information}

\paragraph{Data Availability Statement} The full ODAC23 dataset and all the trained ML models are publicly available at the official ODAC website (\url{https://open-dac.github.io/}) and the open-source GitHub repository (\url{https://github.com/Open-Catalyst-Project/ocp}). All scripts, input, and output files for the calculations presented in this work are available in the supplementary \texttt{deform\_public.zip} file. \texttt{README.md} in \texttt{deform\_public.zip} contains a detailed file structure and description for each directory and file type. A brief outline of the directory is provided below.

\begin{enumerate}
    \item analysis: All data files, figure generation scripts, PNG files for each figure, and Excel spreadsheet containing all raw data used for figures
    \item build\_dataset: Dataset selection scripts and output
    \item example\_mlff\_workflow: Scripts for M3GNet relaxations, which were used for all MLFF relaxations by changing the ASE calculator.
    \item outputs: Input structures, relaxed structures, and energies
    \begin{enumerate}[label=(\alph*)]
        \item DFT: VASP inputs and outputs
        \item UFF: LAMMPS inputs and outputs
        \item M3GNet
        \item CHGNet
        \item MACE (MP-0 and MPA-0)
        \item eSEN
        \item EqV2
    \end{enumerate}
\end{enumerate}
\newpage


\setcounter{figure}{0}
\setcounter{table}{0}
\renewcommand{\thefigure}{S\arabic{figure}}
\renewcommand{\thetable}{S\arabic{table}}

\begin{algorithm}
\caption{Dataset selection algorithm}\label{euclid}

\begin{algorithmic}[1]
\While{len(\texttt{dataset}) $<$ 60}
    \For{\textbf{each} \texttt{sys} \textbf{in} \texttt{IS2RE test set}}
        \State $ \texttt{ads} \gets \texttt{get\_ads}(\texttt{sys}) $
        \State $ \texttt{favor} \gets \texttt{get\_DAC\_favor}(\texttt{sys}) $
        \If{$|\texttt{E\_MOF\_deform}| < 0.05$ \text{eV}}
            \State $ \texttt{class} \gets \text{``Negligible Deformation''} $
        \ElsIf{$|\texttt{E\_MOF\_deform}| \geq 0.05$ \text{eV}}
            \State $ \texttt{class} \gets \text{``Significant Deformation''} $
        \Else
            \State \texttt{skip}
        \EndIf
        \Statex \vspace{0.5em}
        \If{MOF not in \texttt{class} dataset}
            \If{len(\texttt{class} dataset) $<$ 30}
                \If{len(\texttt{sys} with \texttt{ads} in \texttt{class} dataset) $<$ 15}
                    \If{\texttt{favor} is \texttt{very}}
                        \State $ \texttt{dataset}\text{.append(\texttt{sys})} $
                    \ElsIf{\texttt{favor} is \texttt{potential} \textbf{and} no remaining very favorable systems}
                         \State $ \texttt{dataset}\text{.append(\texttt{sys})} $
                    \ElsIf{no remaining potentially favorable systems}
                        \State $ \texttt{dataset}\text{.append(\texttt{sys})} $
                    \EndIf
                \EndIf
            \EndIf
        \EndIf
    \EndFor
\EndWhile
\end{algorithmic}
\end{algorithm}

\begin{figure}[ht!]
    \centering
    \includegraphics[width=\textwidth]{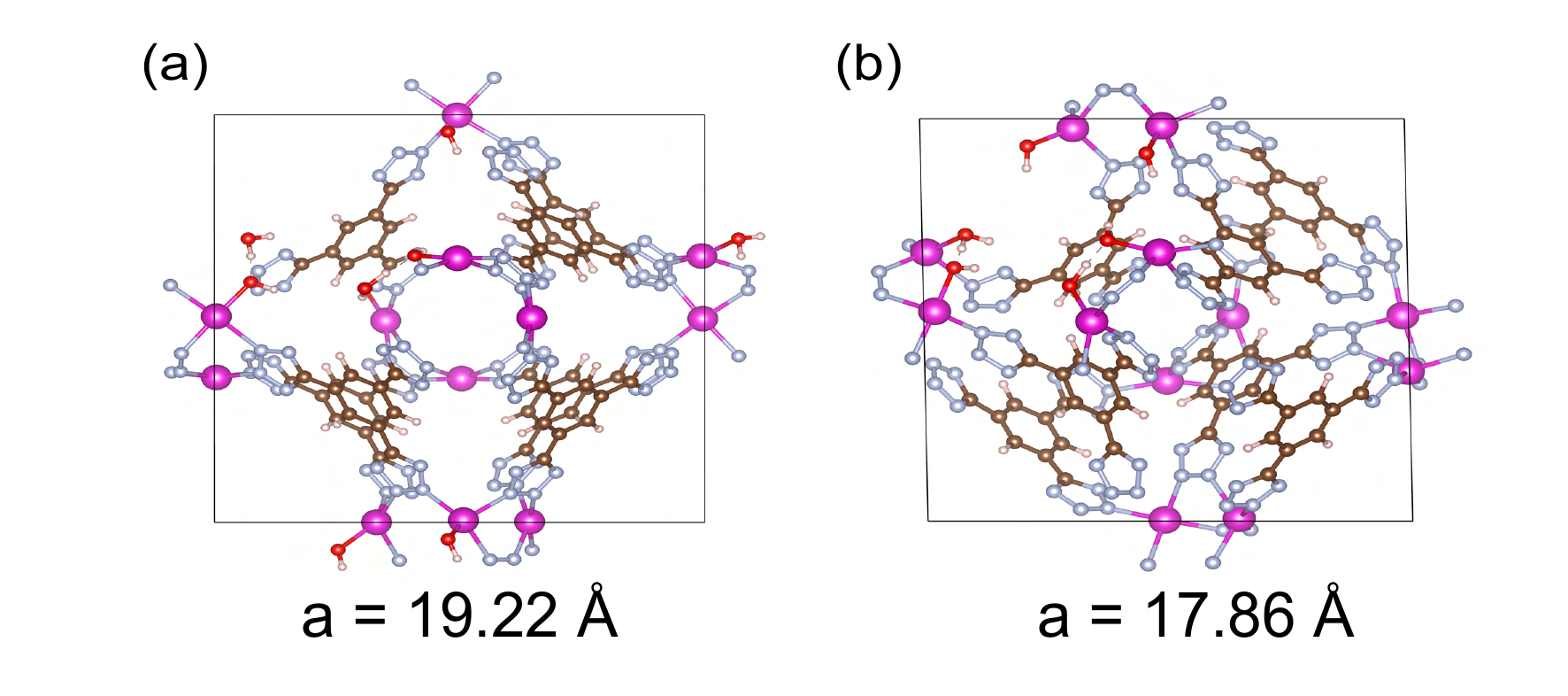}
    \caption{Relaxed empty MOF and corresponding lattice parameter, \textit{a}, for system $3\_78$ (BETHIT\_0.12\_0) from (a) ODAC23 and (b) this work.}
    \label{fig:pore_collapse}
\end{figure}

\begin{longtable}{lrrrrr}
    \caption{Summary of MOF+adsorbate systems considered in this study} \label{tab:systems} \\

    \toprule
    ID & MOF & Defective & Adsorbate & Deformation class & DAC favorability\\
    \midrule
    \endfirsthead
    
    \multicolumn{6}{c}%
    {{\bfseries Table \thetable\ Continued from Previous Page}} \\
    \toprule
    ID & MOF & Defective & Adsorbate & Deformation class & DAC favorability\\
    \midrule
    \endhead
    
    \midrule
    \multicolumn{6}{r}{{Continued on Next Page}} \\
    \endfoot
    
    \bottomrule
    \endlastfoot

    $0\_72^{*}$ & QOPCEE\_0.06\_0 & Yes & \ce{CO2} & Negligible & No \\
    0\_227 & CAYDIR & No & \ce{CO2} & Negligible & No \\
    1\_230 & DAWBUA & No & \ce{CO2} & Negligible & Potential \\
    1\_277 & WORSUT & No & \ce{CO2} & Negligible & Potential \\
    2\_45 & JOHSIJ\_0.07\_1 & Yes & \ce{CO2} & Negligible & Yes \\
    2\_173 & ESURER03 & No & \ce{CO2} & Negligible & No \\
    3\_74 & BETGUE\_0.12\_0 & Yes & \ce{CO2} & Negligible & Potential \\
    3\_88 & IRELAU\_0.12\_0 & Yes & \ce{CO2} & Negligible & Potential \\
    3\_144 & UTEWOG\_0.12\_0 & Yes & \ce{CO2} & Negligible & Yes \\
    $4\_194^{*}$ & FARBIL & No & \ce{CO2} & Negligible & Potential \\
    4\_207 & JIZJOT & No & \ce{CO2} & Negligible & Yes\\
    5\_100 & EGEJIK & No & \ce{CO2} & Negligible & Potential \\
    5\_134 & PIWSEV & No & \ce{CO2} & Negligible & Yes \\
    5\_145 & SIVWUR & No & \ce{CO2} & Negligible & Yes \\
    5\_186 & PIWSEV & No & \ce{CO2} & Negligible & Yes\\
    6\_97 & QUSSII\_0.08\_0 & Yes & \ce{CO2} & Negligible & Potential \\
    7\_91 & MIMFOF\_0.03\_1 & Yes & \ce{CO2} & Negligible & Yes \\
    7\_112 & YARGAB\_0.16\_0 & Yes & \ce{CO2} & Negligible & Potential \\
    7\_302 & ZAFXAI & No & \ce{CO2} & Negligible & Yes \\
    8\_291 & WANJIF & No & \ce{CO2} & Negligible & Yes \\
    8\_292 & WANJIF & No & \ce{CO2} & Negligible & Yes \\
    9\_66 & MALROJ\_0.16\_0 & Yes & \ce{CO2} & Negligible & Yes \\
    \midrule
    0\_419 & LAMGUB & No & \ce{H2O} & Negligible & No \\
    0\_426 & OZAVES & No & \ce{H2O} & Negligible & No \\
    0\_441 & FAHGAY & No & \ce{H2O} & Negligible & No \\
    1\_211 & GUVZII\_0.11\_0 & Yes & \ce{H2O} & Negligible & No \\
    1\_479 & NARPAA & No & \ce{H2O} & Negligible & No \\
    2\_418 & XENMOU\_SL & No & \ce{H2O} & Negligible & No \\
    $2\_432^{*}$ & IRELAU & No & \ce{H2O} & Negligible & No \\
    3\_266 & RIPKIM\_0.08\_1 & Yes & \ce{H2O} & Negligible & No \\
    3\_387 & MABJUV01 & No & \ce{H2O} & Negligible & No \\
    4\_145 & ENISOK\_0.12\_0 & Yes & \ce{H2O} & Negligible & No \\
    4\_148 & KALFOU\_0.16\_0 & Yes & \ce{H2O} & Negligible & Potential \\
    4\_167 & KALFOU\_0.16\_0 & Yes & \ce{H2O} & Negligible & Potential \\
    $4\_494^{*}$ & FARBIL & No & \ce{H2O} & Negligible & Potential \\
    6\_229 & MIMSUY\_0.08\_1 & Yes & \ce{H2O} & Negligible & Yes \\
    $6\_430^{*}$ & EKOPOK & No & \ce{H2O} & Negligible & Potential \\
    8\_203 & CEGDUO\_0.03\_0 & Yes & \ce{H2O} & Negligible & Potential \\
    8\_209 & FAHGAY\_0.12\_0 & Yes & \ce{H2O} & Negligible & Potential \\
    8\_226 & QAVWAN\_0.08\_0 & Yes & \ce{H2O} & Negligible & Potential \\
    9\_188 & KEDJAG14\_0.02\_0 & Yes & \ce{H2O} & Negligible & Yes \\
    9\_444 & VAGMAT & No & \ce{H2O} & Negligible & Potential \\
    \midrule
    2\_74 & JOHSIJ\_0.07\_1 & Yes & \ce{CO2} & Significant & Yes \\
    2\_168 & ECUDOX & No & \ce{CO2} & Significant & Yes \\
    $3\_78^{*,\dag}$ & BETHIT\_0.12\_0 & Yes & \ce{CO2} & Significant & No \\
    6\_70 & KOFPEB04\_0.08\_0 & Yes & \ce{CO2} & Significant & Yes \\
    6\_73 & KOFPEB04\_0.08\_0 & Yes & \ce{CO2} & Significant & Yes \\
    6\_276 & LUSHOX & No & \ce{CO2} & Significant & Yes \\
    7\_94 & MIMFOF\_0.03\_1 & Yes & \ce{CO2} & Significant & Yes \\
    9\_289 & RIPKIM & No & \ce{CO2} & Significant & Yes \\
    \midrule
    0\_217 & WOBHEB\_0.11\_0 & Yes & \ce{H2O} & Significant & No \\
    $0\_452^{*}$ & SIVKAK & No & \ce{H2O} & Significant & No \\
    3\_384 & GOSDEZ & No & \ce{H2O} & Significant & No \\
    $3\_402^{*}$ & OVUCOY & No & \ce{H2O} & Significant & No \\
    4\_469 & LUKREO & No & \ce{H2O} & Significant & No \\
    5\_85 & OZAVES\_0.12\_0 & Yes & \ce{H2O} & Significant & No \\
    $5\_361^{*}$ & DEWRIH & No & \ce{H2O} & Significant & No \\
    6\_210 & EGEJIK\_0.08\_0 & Yes & \ce{H2O} & Significant & Potential \\
    6\_239 & QUSSII\_0.08\_0 & Yes & \ce{H2O} & Significant & Potential \\
    7\_245 & ESEVIH\_0.06\_0 & Yes & \ce{H2O} & Significant & Potential \\

    \bottomrule
\end{longtable}
\vspace{-1em} 
{\footnotesize
\noindent
$^*$MOF geometry initialized from ODAC23 re-relaxation initial positions. \\
$^\dag$Omitted from all analyses due to pore collapse; see Figure~\ref{fig:pore_collapse}.
}

\newpage
\begin{figure}[ht!]
    \centering
    \includegraphics[width=\textwidth]{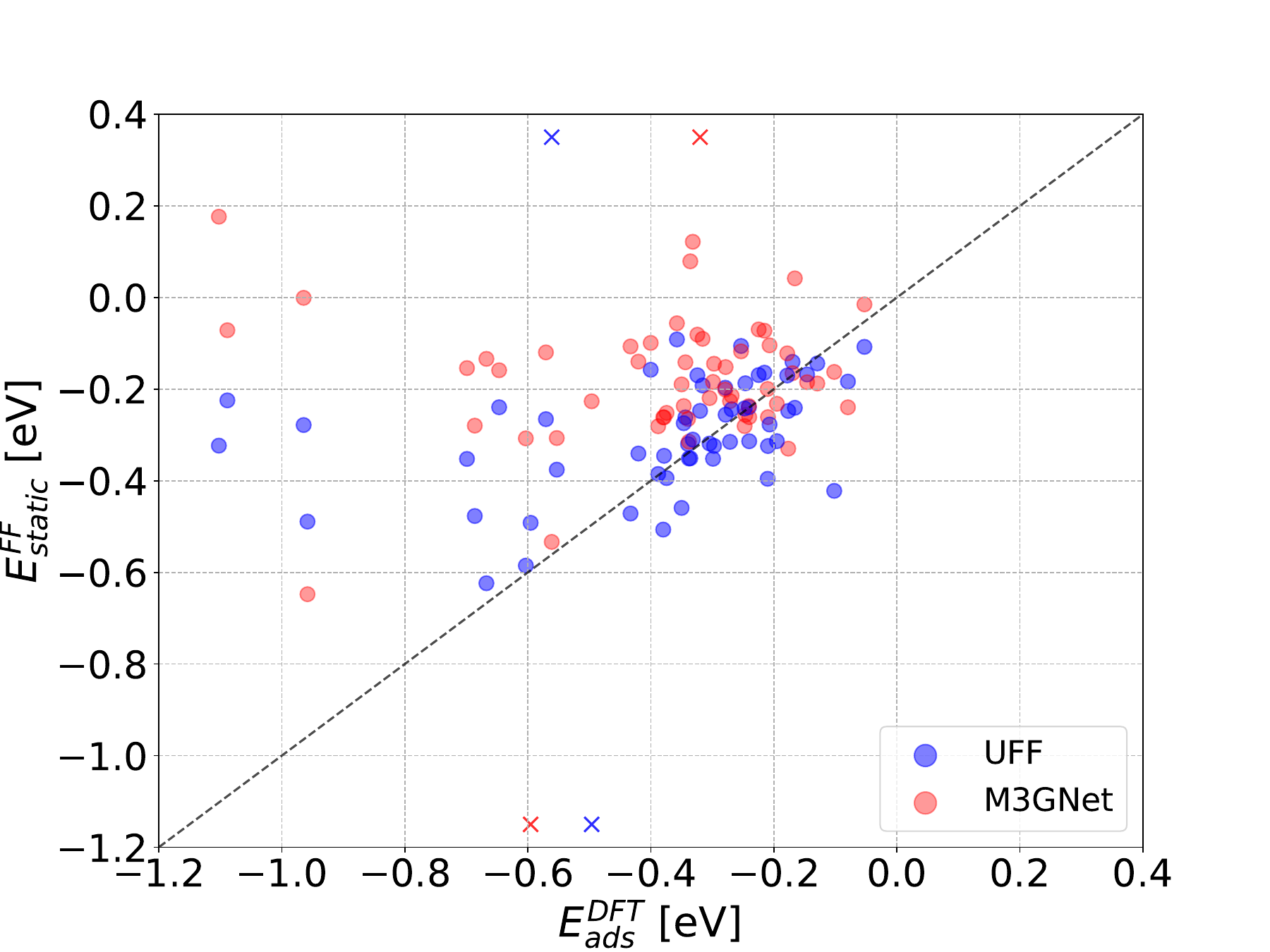}
    \caption{Comparison of DFT $E_{ads}$ with fully flexible MOFs to $E_{static}$ in rigid MOFs using UFF and M3GNet. X markers indicate a point lying beyond the bounds of the plot.}
    \label{fig:rigid_scatter_vs_DFT}
\end{figure}

\newpage
\begin{figure}[ht!]
    \centering
    \includegraphics[width=\textwidth]{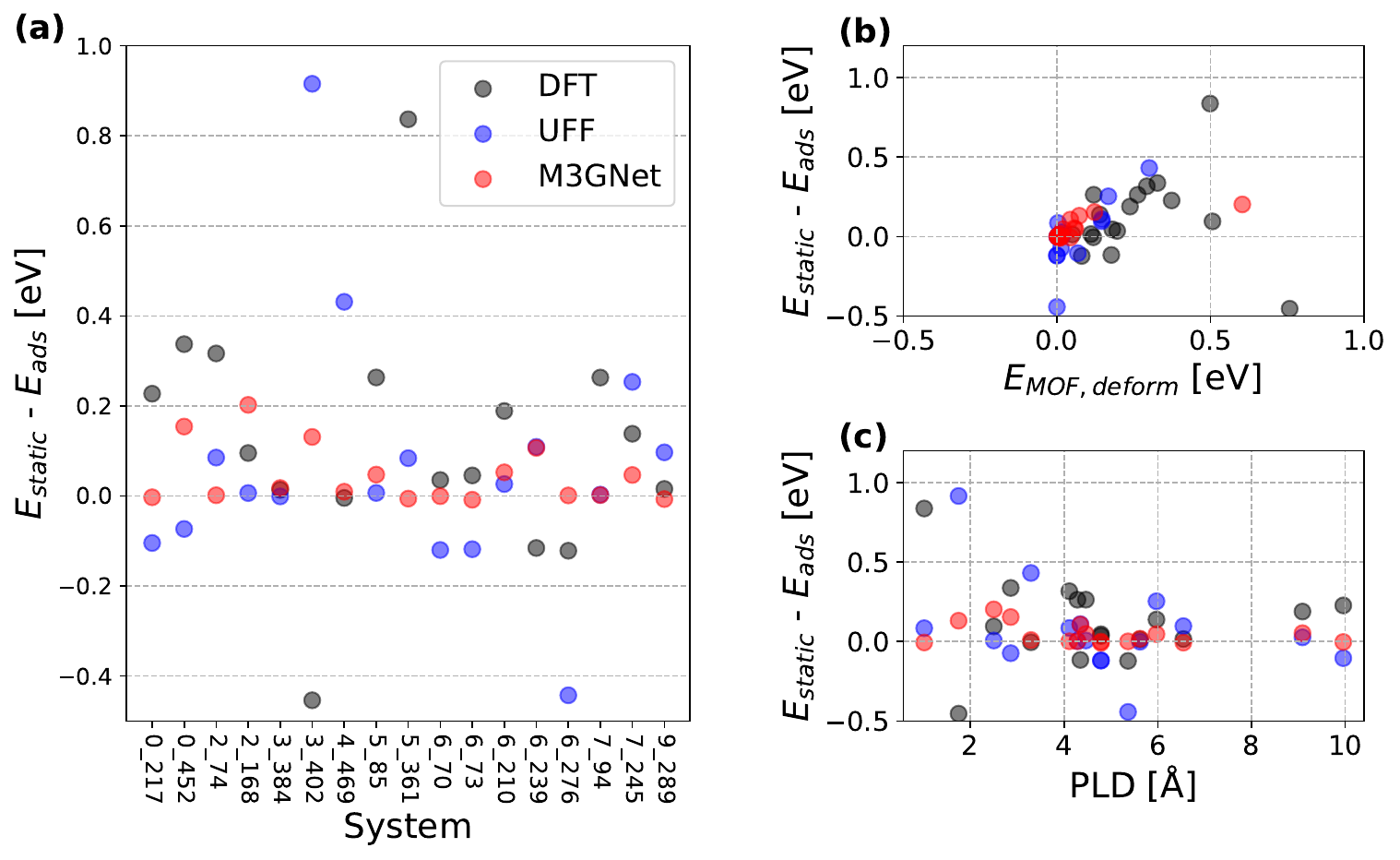}
    \caption{(a) Errors between $E_{static}$ and $E_{ads}$ using DFT, UFF, and M3GNet for 30 systems which undergo significant MOF deformation. (b) Comparison of errors to DFT MOF deformation energies. (c) Comparison of errors to pore limiting diameter (PLD). In all subplots, the y-axis is limited to [-0.5, 1.0] eV for viewing, and points outside this range are omitted.}
    \label{fig:GCMC_by_system}
\end{figure}

\newpage
\begin{figure*}[ht!]
    \centering
    \includegraphics[angle=90, width=\textwidth, height=0.9\textheight, keepaspectratio]{figures/combined_no_deform.pdf}
    \caption{Adsorption energy errors (upper left triangle), interaction energy errors (bottom triangle), and MOF deformation errors (right triangle) from FF relaxations for systems that undergo negligible deformation separated by adsorbate. All energies are in eV using the color bar shown in the figure.}
    \label{fig:gridplot_supp}
\end{figure*}

\newpage
\begin{figure}[ht!]
    \centering
    \includegraphics[width=\textwidth]{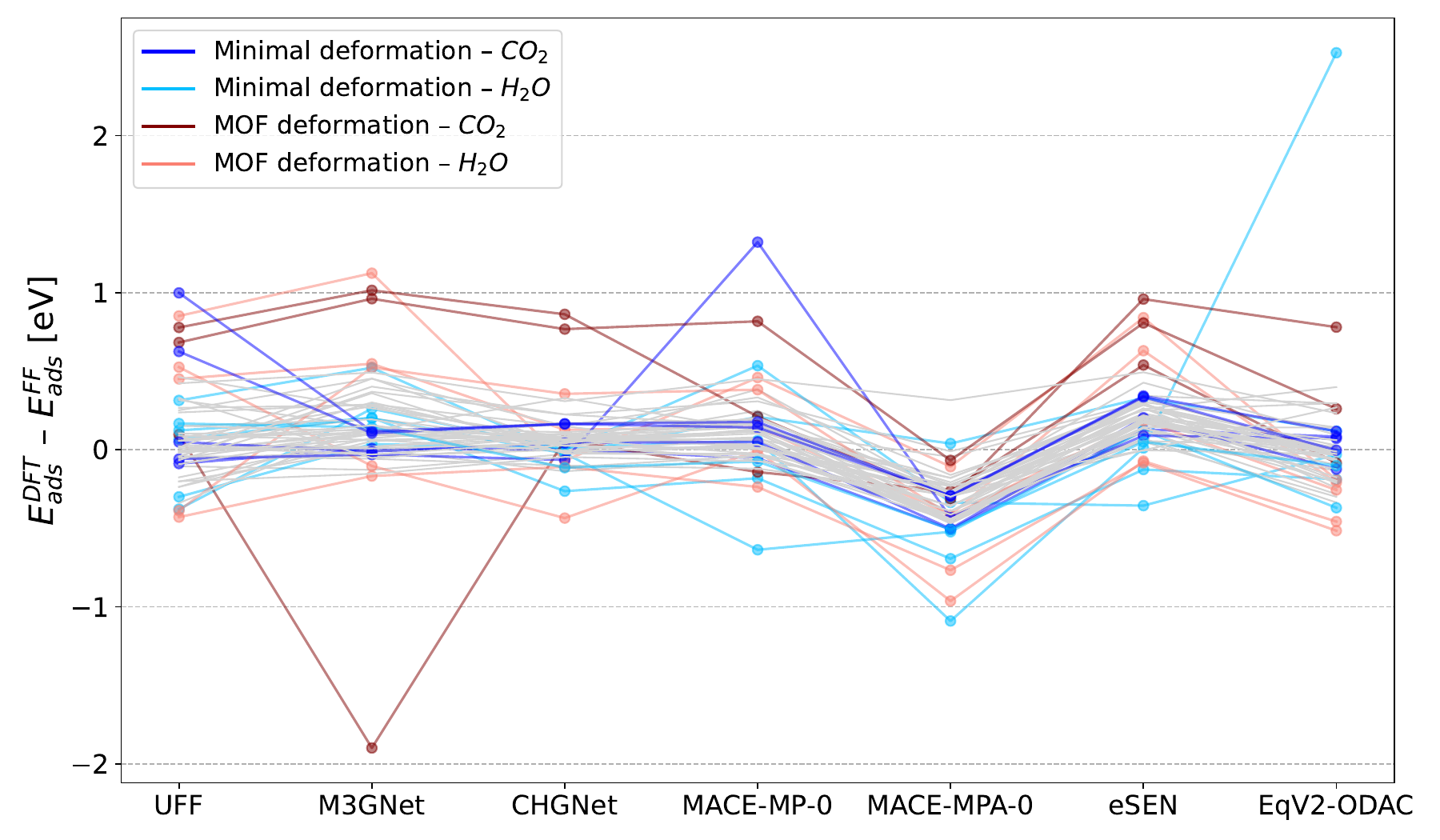}
    \caption{Adsorption energy errors where errors for one MOF+adsorbate system are connected by lines to aid visualization. Systems with errors less than 0.2 eV for all seven FFs are excluded for simplicity. Plots are bound on the y-axis from [-2,2.5] eV for scaling purposes.}
    \label{fig:err_by_system}
\end{figure}

\newpage
\begin{figure}[ht!]
    \centering
    \includegraphics[width=\textwidth]{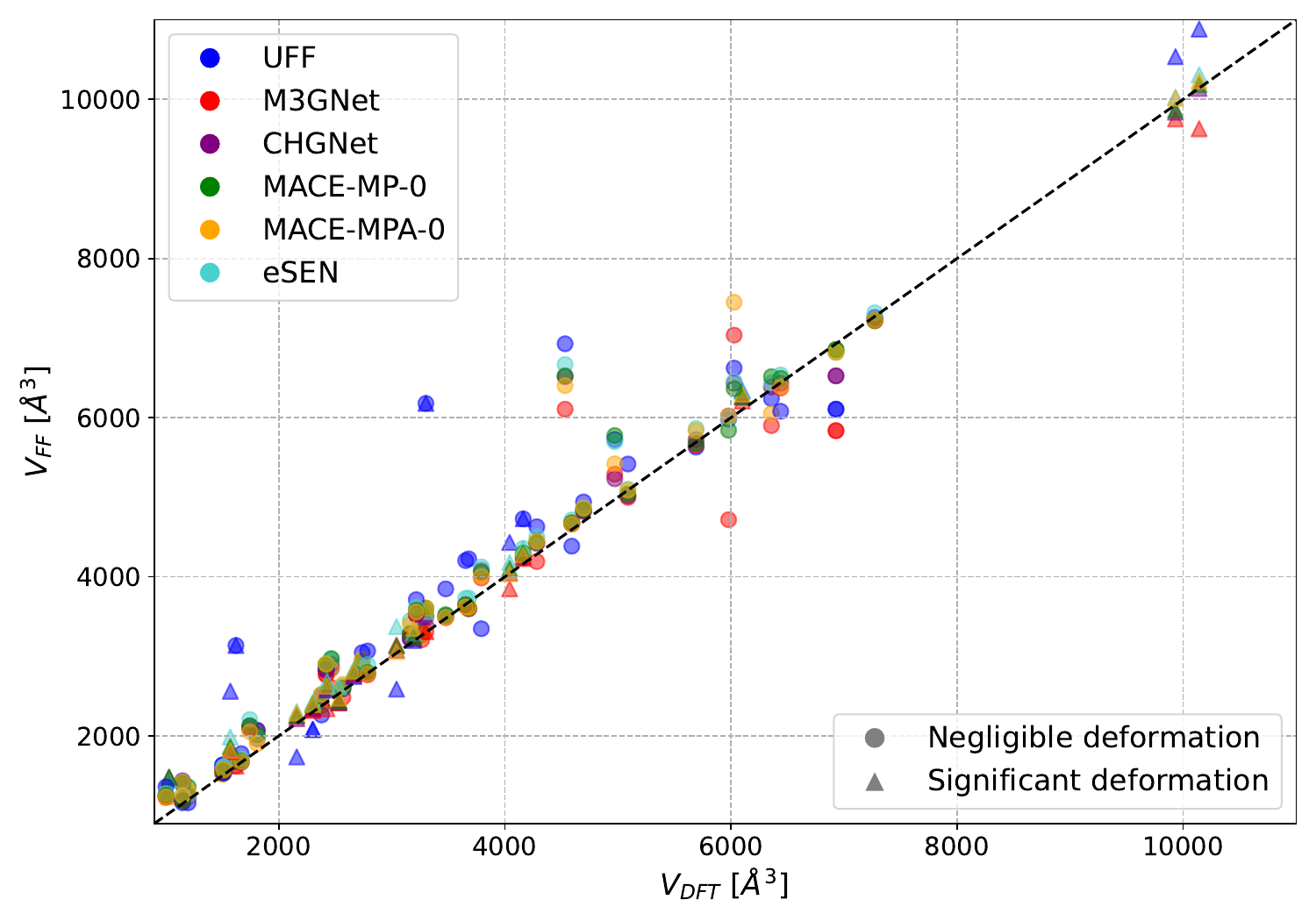}
    \caption{Parity plot of volumes determined by DFT and FF relaxation. Colors denote the FF used for the y-axis, and shapes denote whether the system undergoes significant deformation according to DFT.}
    \label{fig:volumes}
\end{figure}

\FloatBarrier
\newpage
\begin{table}
    \centering
    \setlength{\tabcolsep}{5pt}
    \resizebox{0.4\linewidth}{!}{
        \begin{tabular}{lcc}
            \toprule
            FF & $R^{2}$ & \textit{r}\\
            \midrule
            UFF & –0.783 & 0.235 \\
            M3GNet & –2.062 & 0.152 \\
            CHGNet & 0.248 & 0.648 \\
            MACE-MP-0 & –0.507 & 0.549 \\
            MACE-MPA-0 & –2.044 & 0.631 \\
            eSEN & -0.786 & 0.477 \\
            EqV2-ODAC & –1.838 & 0.540 \\
            \bottomrule
        \end{tabular}
    }
    \caption{$R^{2}$ and Pearson correlation coefficients (\textit{r}) for FF $E_{ads}$ predictions relative to DFT. The scikit-learn implementation for $R^{2}$ was used as defined by Eqn. \ref{eq:R2}. $R^{2}$ is negative when the FF predictions are worse than predicting the mean of the DFT adsorption energies}
    \label{tab:R2}
\end{table}

\begin{figure}[ht!]
    \centering
    \includegraphics[width=\textwidth]{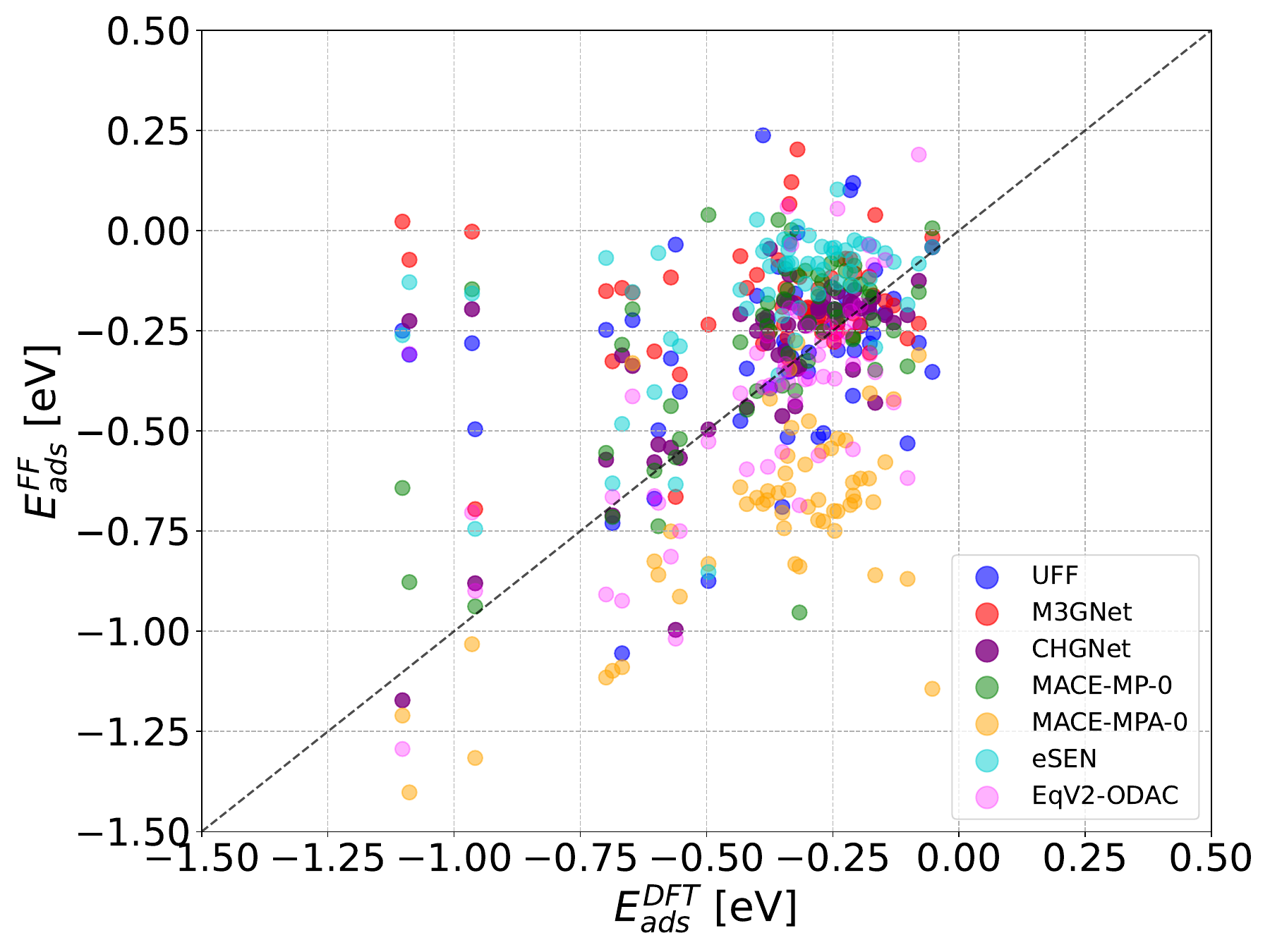}
    \caption{Comparison of DFT $E_{ads}$ with FF $E_{ads}$ predictions for all models tested. Points falling outside the [-1.5, 0.5] eV range are excluded.}
    \label{fig:parity}
\end{figure}

\begin{table*}
    \centering
    \setlength{\tabcolsep}{5pt}
    \resizebox{\textwidth}{!}{
        \begin{tabular}{lcccc}
            \toprule
            System & \shortstack{Shortest distance between\\ MOF and adsorbate [Å]} & \shortstack{UFF $E_{ads}$\\ error [eV]} & \shortstack{CHGNet $E_{ads}$\\ error [eV]} & \shortstack{MACE-MP-0 $E_{ads}$\\ error [eV]}\\
            \midrule
            2\_74 & 1.926 & 0.779 & 0.863 & 0.211  \\
            4\_469 & 1.608 & –0.388 & 0.356 & 0.382  \\
            7\_94 & 1.970 & 0.683 & 0.768 & 0.818 \\
            \bottomrule
        \end{tabular}
    }
    \caption{FF errors for select systems with short distances between MOF and adsorbate in the DFT-relaxed geometry}
    \label{tab:short_errors}
\end{table*}

\begin{table*}
    \centering
    \setlength{\tabcolsep}{5pt}
    \resizebox{\textwidth}{!}{
        \begin{tabular}{lccccc}
            \toprule
            FF & \shortstack{\ce{CO2} $E_{adsorbate}$ \\ correction [eV]} & \shortstack{\ce{H2O} $E_{adsorbate}$ \\ correction [eV]} & \shortstack{MAE with direct \\ $E_{adsorbate}$ [eV]} & \shortstack{MAE with corrected \\ $E_{adsorbate}$ [eV]} & \shortstack{$R^{2}$ with corrected \\ $E_{adsorbate}$}\\
            \midrule
            UFF & –0.102 & –0.077 & 0.213 & 0.221 & –0.632 \\
            M3GNet & –0.064 & –0.245 & 0.247 & 0.196 & –1.458 \\
            CHGNet & –0.135 & –0.006 & 0.124 & 0.113 & 0.414 \\
            MACE-MP-0 & –0.176 & –0.068 & 0.181 & 0.160 & –0.180 \\
            MACE-MPA-0 & 0.305 & 0.385 & 0.357 & 0.139 & 0.200 \\
            eSEN & -0.271 & -0.174 & 0.243 & 0.143 & 0.169 \\
            \bottomrule
        \end{tabular}
    }
    \caption{$E_{ads}$ Error statistics for MLFFs using corrected gas-phase adsorbate energies ($E_{adsorbate}$)}
    \label{tab:gas_errors}
\end{table*}

\subsection{Methods for gas-phase adsorbate energy corrections}
To validate that gas-phase errors did not significantly affect conclusions presented in the main text, we used a regression approach to determine gas-phase corrections as shown in Table \ref{tab:gas_errors}. For each FF in Table \ref{tab:gas_errors}, linear regression was used to compute a correction to the gas-phase \ce{CO2} and \ce{H2O} molecular energies such that the errors between DFT $E_{ads}$ and FF $E_{ads}$ were minimized. The regression was done using the scikit-learn Python package.\cite{sklearn} The regression problem is formally defined as
\begin{align}
d = x + \mathbf{M}b
\label{eq:gas_corr}
\end{align}
where $d$ is a $59\times1$ vector containing the DFT-calculated $E_{ads}$ for each system, $x$ is a $59\times1$ vector of FF-calculated $E_{ads}$, $\mathbf{M}$ is a $59\times2$ matrix encoding whether a system contains \ce{CO2} or \ce{H2O}, and $b$ is a $2\times1$ vector of corrections to be computed by the regression. The feature matrix $\mathbf{M}$ contains only 0 and 1, with a 1 in the first column indicating a \ce{CO2} system and a 1 in the second column indicating a \ce{H2O} system. The other column is always 0. This in effect creates two distinct linear regression problems, resulting in one correction for \ce{CO2} that minimizes the error between DFT and FF for the \ce{CO2} systems, and likewise for \ce{H2O}. A separate regression problem was solved for each FF, so each pair of corrections is different as shown in Table \ref{tab:gas_errors}.


\begin{mcitethebibliography}{87}
\providecommand*\natexlab[1]{#1}
\providecommand*\mciteSetBstSublistMode[1]{}
\providecommand*\mciteSetBstMaxWidthForm[2]{}
\providecommand*\mciteBstWouldAddEndPuncttrue
  {\def\EndOfBibitem{\unskip.}}
\providecommand*\mciteBstWouldAddEndPunctfalse
  {\let\EndOfBibitem\relax}
\providecommand*\mciteSetBstMidEndSepPunct[3]{}
\providecommand*\mciteSetBstSublistLabelBeginEnd[3]{}
\providecommand*\EndOfBibitem{}
\mciteSetBstSublistMode{f}
\mciteSetBstMaxWidthForm{subitem}{(\alph{mcitesubitemcount})}
\mciteSetBstSublistLabelBeginEnd
  {\mcitemaxwidthsubitemform\space}
  {\relax}
  {\relax}

\bibitem[Sholl and Lively(2016)Sholl, and Lively]{SevenSep}
Sholl,~D.~S.; Lively,~R.~P. Seven chemical separations to change the world. \emph{Nature} \textbf{2016}, \emph{532}, 435--437\relax
\mciteBstWouldAddEndPuncttrue
\mciteSetBstMidEndSepPunct{\mcitedefaultmidpunct}
{\mcitedefaultendpunct}{\mcitedefaultseppunct}\relax
\EndOfBibitem
\bibitem[Beuttler \latin{et~al.}(2019)Beuttler, Charles, and Wurzbacher]{Beuttler2019}
Beuttler,~C.; Charles,~L.; Wurzbacher,~J. The Role of Direct Air Capture in Mitigation of Anthropogenic Greenhouse Gas Emissions. \emph{Frontiers in Climate} \textbf{2019}, \emph{1}\relax
\mciteBstWouldAddEndPuncttrue
\mciteSetBstMidEndSepPunct{\mcitedefaultmidpunct}
{\mcitedefaultendpunct}{\mcitedefaultseppunct}\relax
\EndOfBibitem
\bibitem[Sanz-Pérez \latin{et~al.}(2016)Sanz-Pérez, Murdock, Didas, and Jones]{DAC_review}
Sanz-Pérez,~E.~S.; Murdock,~C.~R.; Didas,~S.~A.; Jones,~C.~W. Direct Capture of CO$_{2}$ from Ambient Air. \emph{Chemical Reviews} \textbf{2016}, \emph{116}, 11840--11876, PMID: 27560307\relax
\mciteBstWouldAddEndPuncttrue
\mciteSetBstMidEndSepPunct{\mcitedefaultmidpunct}
{\mcitedefaultendpunct}{\mcitedefaultseppunct}\relax
\EndOfBibitem
\bibitem[Kim \latin{et~al.}(2022)Kim, Landa, Ravutla, Realff, and Boukouvala]{Kim2022}
Kim,~S.~H.; Landa,~H. O.~R.; Ravutla,~S.; Realff,~M.~J.; Boukouvala,~F. Data-driven simultaneous process optimization and adsorbent selection for vacuum pressure swing adsorption. \emph{Chemical Engineering Research and Design} \textbf{2022}, \emph{188}, 1013--1028\relax
\mciteBstWouldAddEndPuncttrue
\mciteSetBstMidEndSepPunct{\mcitedefaultmidpunct}
{\mcitedefaultendpunct}{\mcitedefaultseppunct}\relax
\EndOfBibitem
\bibitem[Furukawa \latin{et~al.}(2013)Furukawa, Cordova, O’Keeffe, and Yaghi]{Furukawa2013}
Furukawa,~H.; Cordova,~K.~E.; O’Keeffe,~M.; Yaghi,~O.~M. The Chemistry and Applications of Metal-Organic Frameworks. \emph{Science} \textbf{2013}, \emph{341}, 1230444\relax
\mciteBstWouldAddEndPuncttrue
\mciteSetBstMidEndSepPunct{\mcitedefaultmidpunct}
{\mcitedefaultendpunct}{\mcitedefaultseppunct}\relax
\EndOfBibitem
\bibitem[Moosavi \latin{et~al.}(2020)Moosavi, Nandy, Jablonka, Ongari, Janet, Boyd, Lee, Smit, and Kulik]{Moosavi2020}
Moosavi,~S.~M.; Nandy,~A.; Jablonka,~K.~M.; Ongari,~D.; Janet,~J.~P.; Boyd,~P.~G.; Lee,~Y.; Smit,~B.; Kulik,~H.~J. Understanding the diversity of the metal-organic framework ecosystem. \emph{Nature Communications} \textbf{2020}, \emph{11}\relax
\mciteBstWouldAddEndPuncttrue
\mciteSetBstMidEndSepPunct{\mcitedefaultmidpunct}
{\mcitedefaultendpunct}{\mcitedefaultseppunct}\relax
\EndOfBibitem
\bibitem[Deria \latin{et~al.}(2015)Deria, Li, Zhang, Snurr, Hupp, and Farha]{Deria2015}
Deria,~P.; Li,~S.; Zhang,~H.; Snurr,~R.~Q.; Hupp,~J.~T.; Farha,~O.~K. A MOF platform for incorporation of complementary organic motifs for CO$_{2}$ binding. \emph{Chem. Commun.} \textbf{2015}, \emph{51}, 12478--12481\relax
\mciteBstWouldAddEndPuncttrue
\mciteSetBstMidEndSepPunct{\mcitedefaultmidpunct}
{\mcitedefaultendpunct}{\mcitedefaultseppunct}\relax
\EndOfBibitem
\bibitem[Yang \latin{et~al.}(2024)Yang, Shin, Ooe, Hasegawa, Yamane, Yamada, van Duin, Murase, and Mauro]{Yang2024}
Yang,~Y.; Shin,~Y.~K.; Ooe,~H.; Hasegawa,~U.; Yamane,~S.; Yamada,~H.; van Duin,~A.~C.; Murase,~Y.; Mauro,~J.~C. Adsorption of CO$_{2}$ by Amine-Functionalized Metal–Organic Frameworks Using GCMC and ReaxFF-Based Metadynamics Simulations. \emph{The Journal of Physical Chemistry C} \textbf{2024}, \emph{128}, 5257--5270\relax
\mciteBstWouldAddEndPuncttrue
\mciteSetBstMidEndSepPunct{\mcitedefaultmidpunct}
{\mcitedefaultendpunct}{\mcitedefaultseppunct}\relax
\EndOfBibitem
\bibitem[Tiainen \latin{et~al.}(2022)Tiainen, Mannisto, Tenhu, and Hietala]{Tiainen2022}
Tiainen,~T.; Mannisto,~J.~K.; Tenhu,~H.; Hietala,~S. CO$_{2}$ Capture and Low-Temperature Release by Poly(aminoethyl methacrylate) and Derivatives. \emph{Langmuir} \textbf{2022}, \emph{38}, 5197--5208, PMID: 34879650\relax
\mciteBstWouldAddEndPuncttrue
\mciteSetBstMidEndSepPunct{\mcitedefaultmidpunct}
{\mcitedefaultendpunct}{\mcitedefaultseppunct}\relax
\EndOfBibitem
\bibitem[Senkovska \latin{et~al.}(2023)Senkovska, Bon, Abylgazina, Mendt, Berger, Kieslich, Petkov, Luiz~Fiorio, Joswig, Heine, Schaper, Bachetzky, Schmid, Fischer, Pöppl, Brunner, and Kaskel]{Senkovska2023}
Senkovska,~I. \latin{et~al.}  Understanding MOF Flexibility: An Analysis Focused on Pillared Layer MOFs as a Model System. \emph{Angewandte Chemie International Edition} \textbf{2023}, \emph{62}, e202218076\relax
\mciteBstWouldAddEndPuncttrue
\mciteSetBstMidEndSepPunct{\mcitedefaultmidpunct}
{\mcitedefaultendpunct}{\mcitedefaultseppunct}\relax
\EndOfBibitem
\bibitem[Agrawal and Sholl(2019)Agrawal, and Sholl]{Agrawal2019}
Agrawal,~M.; Sholl,~D.~S. Effects of Intrinsic Flexibility on Adsorption Properties of Metal–Organic Frameworks at Dilute and Nondilute Loadings. \emph{ACS Applied Materials \& Interfaces} \textbf{2019}, \emph{11}, 31060--31068, PMID: 31333011\relax
\mciteBstWouldAddEndPuncttrue
\mciteSetBstMidEndSepPunct{\mcitedefaultmidpunct}
{\mcitedefaultendpunct}{\mcitedefaultseppunct}\relax
\EndOfBibitem
\bibitem[Han \latin{et~al.}(2020)Han, Yang, and Sholl]{Han2020}
Han,~C.; Yang,~Y.; Sholl,~D.~S. Quantitatively Predicting Impact of Structural Flexibility on Molecular Diffusion in Small Pore Metal–Organic Frameworks—A Molecular Dynamics Study of Hypothetical ZIF-8 Polymorphs. \emph{The Journal of Physical Chemistry C} \textbf{2020}, \emph{124}, 20203--20212\relax
\mciteBstWouldAddEndPuncttrue
\mciteSetBstMidEndSepPunct{\mcitedefaultmidpunct}
{\mcitedefaultendpunct}{\mcitedefaultseppunct}\relax
\EndOfBibitem
\bibitem[Sholl \latin{et~al.}(2021)Sholl, Daou, Findley, Fang, Boulfelfel, and Ravikovitch]{Daou2021}
Sholl,~D.~S.; Daou,~A.~S.; Findley,~J.~M.; Fang,~H.; Boulfelfel,~S.~E.; Ravikovitch,~P.~I. {Quantifying impact of intrinsic flexibility on molecular adsorption in zeolites}. \emph{Journal of Physical Chemistry C} \textbf{2021}, \emph{125}, 5296--5305\relax
\mciteBstWouldAddEndPuncttrue
\mciteSetBstMidEndSepPunct{\mcitedefaultmidpunct}
{\mcitedefaultendpunct}{\mcitedefaultseppunct}\relax
\EndOfBibitem
\bibitem[Serre \latin{et~al.}(2007)Serre, Mellot-Draznieks, Surblé, Audebrand, Filinchuk, and Férey]{Serre2007}
Serre,~C.; Mellot-Draznieks,~C.; Surblé,~S.; Audebrand,~N.; Filinchuk,~Y.; Férey,~G. Role of Solvent-Host Interactions That Lead to Very Large Swelling of Hybrid Frameworks. \emph{Science} \textbf{2007}, \emph{315}, 1828--1831\relax
\mciteBstWouldAddEndPuncttrue
\mciteSetBstMidEndSepPunct{\mcitedefaultmidpunct}
{\mcitedefaultendpunct}{\mcitedefaultseppunct}\relax
\EndOfBibitem
\bibitem[Mason \latin{et~al.}(2015)Mason, Oktawiec, Taylor, Hudson, Rodriguez, Bachman, Gonzalez, Cervellino, Guagliardi, Brown, Llewellyn, Masciocchi, and Long]{Mason2015}
Mason,~J.~A.; Oktawiec,~J.; Taylor,~M.~K.; Hudson,~M.~R.; Rodriguez,~J.; Bachman,~J.~E.; Gonzalez,~M.~I.; Cervellino,~A.; Guagliardi,~A.; Brown,~C.~M.; Llewellyn,~P.~L.; Masciocchi,~N.; Long,~J.~R. {Methane storage in flexible metal-organic frameworks with intrinsic thermal management}. \emph{Nature} \textbf{2015}, \emph{527}, 357--361\relax
\mciteBstWouldAddEndPuncttrue
\mciteSetBstMidEndSepPunct{\mcitedefaultmidpunct}
{\mcitedefaultendpunct}{\mcitedefaultseppunct}\relax
\EndOfBibitem
\bibitem[Jawahery \latin{et~al.}(2017)Jawahery, Simon, Braun, Witman, Tiana, Vlaisavljevich, and Smit]{Jawahery2017}
Jawahery,~S.; Simon,~C.~M.; Braun,~E.; Witman,~M.; Tiana,~D.; Vlaisavljevich,~B.; Smit,~B. Adsorbate-induced lattice deformation in IRMOF-74 series. \emph{Nature Communications} \textbf{2017}, \emph{8}\relax
\mciteBstWouldAddEndPuncttrue
\mciteSetBstMidEndSepPunct{\mcitedefaultmidpunct}
{\mcitedefaultendpunct}{\mcitedefaultseppunct}\relax
\EndOfBibitem
\bibitem[Dundar \latin{et~al.}(2017)Dundar, Chanut1, Formalik, Boulet, Llewellyn, and Kuchta]{Dundar2017}
Dundar,~E.; Chanut1,~N.; Formalik,~F.; Boulet,~P.; Llewellyn,~P.~L.; Kuchta,~B. Modeling of adsorption of CO$_{2}$ in the deformed pores of MIL-53(Al). \emph{J. Mol. Model.} \textbf{2017}, \emph{23}\relax
\mciteBstWouldAddEndPuncttrue
\mciteSetBstMidEndSepPunct{\mcitedefaultmidpunct}
{\mcitedefaultendpunct}{\mcitedefaultseppunct}\relax
\EndOfBibitem
\bibitem[Daglar and Keskin(2020)Daglar, and Keskin]{Daglar2020}
Daglar,~H.; Keskin,~S. {Recent advances, opportunities, and challenges in high-throughput computational screening of MOFs for gas separations}. \textbf{2020}, \emph{422}\relax
\mciteBstWouldAddEndPuncttrue
\mciteSetBstMidEndSepPunct{\mcitedefaultmidpunct}
{\mcitedefaultendpunct}{\mcitedefaultseppunct}\relax
\EndOfBibitem
\bibitem[Yang and Sholl(2022)Yang, and Sholl]{Yang2022}
Yang,~Y.; Sholl,~D.~S. {A systematic examination of the impacts of MOF flexibility on intracrystalline molecular diffusivities}. \emph{Journal of Materials Chemistry A} \textbf{2022}, \emph{10}, 4242--4253\relax
\mciteBstWouldAddEndPuncttrue
\mciteSetBstMidEndSepPunct{\mcitedefaultmidpunct}
{\mcitedefaultendpunct}{\mcitedefaultseppunct}\relax
\EndOfBibitem
\bibitem[Oliveira \latin{et~al.}(2023)Oliveira, Cleeton, Ferreira, Luan, Farmahini, Sarkisov, and Steiner]{Oliveira2023}
Oliveira,~F.~L.; Cleeton,~C.; Ferreira,~R. N.~B.; Luan,~B.; Farmahini,~A.~H.; Sarkisov,~L.; Steiner,~M. CRAFTED: An exploratory database of simulated adsorption isotherms of metal-organic frameworks. \emph{Scientific Data} \textbf{2023}, \emph{10}\relax
\mciteBstWouldAddEndPuncttrue
\mciteSetBstMidEndSepPunct{\mcitedefaultmidpunct}
{\mcitedefaultendpunct}{\mcitedefaultseppunct}\relax
\EndOfBibitem
\bibitem[Boyd \latin{et~al.}(2017)Boyd, Moosavi, Witman, and Smit]{Boyd2017}
Boyd,~P.~G.; Moosavi,~S.~M.; Witman,~M.; Smit,~B. Force-Field Prediction of Materials Properties in Metal-Organic Frameworks. \emph{The Journal of Physical Chemistry Letters} \textbf{2017}, \emph{8}, 357--363, PMID: 28008758\relax
\mciteBstWouldAddEndPuncttrue
\mciteSetBstMidEndSepPunct{\mcitedefaultmidpunct}
{\mcitedefaultendpunct}{\mcitedefaultseppunct}\relax
\EndOfBibitem
\bibitem[Yu \latin{et~al.}(2021)Yu, Anstine, Boulfelfel, Gu, Colina, and Sholl]{ZhenziYu2021}
Yu,~Z.; Anstine,~D.~M.; Boulfelfel,~S.~E.; Gu,~C.; Colina,~C.~M.; Sholl,~D.~S. Incorporating Flexibility Effects into Metal–Organic Framework Adsorption Simulations Using Different Models. \emph{ACS Applied Materials \& Interfaces} \textbf{2021}, \emph{13}, 61305--61315\relax
\mciteBstWouldAddEndPuncttrue
\mciteSetBstMidEndSepPunct{\mcitedefaultmidpunct}
{\mcitedefaultendpunct}{\mcitedefaultseppunct}\relax
\EndOfBibitem
\bibitem[Rogge \latin{et~al.}(2019)Rogge, Goeminne, Demuynck, Guti{\'{e}}rrez-Sevillano, Vandenbrande, Vanduyfhuys, Waroquier, Verstraelen, and Van~Speybroeck]{Rogge2019}
Rogge,~S.~M.; Goeminne,~R.; Demuynck,~R.; Guti{\'{e}}rrez-Sevillano,~J.~J.; Vandenbrande,~S.; Vanduyfhuys,~L.; Waroquier,~M.; Verstraelen,~T.; Van~Speybroeck,~V. {Modeling Gas Adsorption in Flexible Metal–Organic Frameworks via Hybrid Monte Carlo/Molecular Dynamics Schemes}. \emph{Advanced Theory and Simulations} \textbf{2019}, \emph{2}\relax
\mciteBstWouldAddEndPuncttrue
\mciteSetBstMidEndSepPunct{\mcitedefaultmidpunct}
{\mcitedefaultendpunct}{\mcitedefaultseppunct}\relax
\EndOfBibitem
\bibitem[Witman \latin{et~al.}(2017)Witman, Ling, Jawahery, Boyd, Haranczyk, Slater, and Smit]{Witman2017}
Witman,~M.; Ling,~S.; Jawahery,~S.; Boyd,~P.~G.; Haranczyk,~M.; Slater,~B.; Smit,~B. {The Influence of Intrinsic Framework Flexibility on Adsorption in Nanoporous Materials}. \emph{Journal of the American Chemical Society} \textbf{2017}, \emph{139}, 5547--5557\relax
\mciteBstWouldAddEndPuncttrue
\mciteSetBstMidEndSepPunct{\mcitedefaultmidpunct}
{\mcitedefaultendpunct}{\mcitedefaultseppunct}\relax
\EndOfBibitem
\bibitem[Coudert \latin{et~al.}(2011)Coudert, Boutin, Jeffroy, Mellot-Draznieks, and Fuchs]{Coudert2011}
Coudert,~F.-X.; Boutin,~A.; Jeffroy,~M.; Mellot-Draznieks,~C.; Fuchs,~A.~H. Thermodynamic Methods and Models to Study Flexible Metal–Organic Frameworks. \emph{ChemPhysChem} \textbf{2011}, \emph{12}, 247--258\relax
\mciteBstWouldAddEndPuncttrue
\mciteSetBstMidEndSepPunct{\mcitedefaultmidpunct}
{\mcitedefaultendpunct}{\mcitedefaultseppunct}\relax
\EndOfBibitem
\bibitem[Sriram \latin{et~al.}(2024)Sriram, Choi, Yu, Brabson, Das, Ulissi, Uyttendaele, Medford, and Sholl]{ODAC23}
Sriram,~A.; Choi,~S.; Yu,~X.; Brabson,~L.~M.; Das,~A.; Ulissi,~Z.; Uyttendaele,~M.; Medford,~A.~J.; Sholl,~D.~S. The Open DAC 2023 Dataset and Challenges for Sorbent Discovery in Direct Air Capture. \emph{ACS Central Science} \textbf{2024}, \emph{10}, 923--941\relax
\mciteBstWouldAddEndPuncttrue
\mciteSetBstMidEndSepPunct{\mcitedefaultmidpunct}
{\mcitedefaultendpunct}{\mcitedefaultseppunct}\relax
\EndOfBibitem
\bibitem[Lee \latin{et~al.}(2021)Lee, Kim, Cho, Lee, Lee, Cho, and Kim]{Lee2021}
Lee,~S.; Kim,~B.; Cho,~H.; Lee,~H.; Lee,~S.~Y.; Cho,~E.~S.; Kim,~J. {Computational Screening of Trillions of Metal-Organic Frameworks for High-Performance Methane Storage}. \emph{ACS Applied Materials and Interfaces} \textbf{2021}, \emph{13}, 23647--23654\relax
\mciteBstWouldAddEndPuncttrue
\mciteSetBstMidEndSepPunct{\mcitedefaultmidpunct}
{\mcitedefaultendpunct}{\mcitedefaultseppunct}\relax
\EndOfBibitem
\bibitem[Nazarian \latin{et~al.}(2017)Nazarian, Camp, Chung, Snurr, and Sholl]{Nazarian2017}
Nazarian,~D.; Camp,~J.~S.; Chung,~Y.~G.; Snurr,~R.~Q.; Sholl,~D.~S. {Large-Scale Refinement of Metal-Organic Framework Structures Using Density Functional Theory}. \emph{Chemistry of Materials} \textbf{2017}, \emph{29}, 2521--2528\relax
\mciteBstWouldAddEndPuncttrue
\mciteSetBstMidEndSepPunct{\mcitedefaultmidpunct}
{\mcitedefaultendpunct}{\mcitedefaultseppunct}\relax
\EndOfBibitem
\bibitem[Li \latin{et~al.}(2022)Li, Cai, Wang, and Wu]{Li2022}
Li,~M.; Cai,~W.; Wang,~C.; Wu,~X. {High-throughput computational screening of hypothetical metal-organic frameworks with open copper sites for CO$_{2}$/H$_{2}$ separation}. \emph{Physical Chemistry Chemical Physics} \textbf{2022}, \relax
\mciteBstWouldAddEndPunctfalse
\mciteSetBstMidEndSepPunct{\mcitedefaultmidpunct}
{}{\mcitedefaultseppunct}\relax
\EndOfBibitem
\bibitem[Qiao \latin{et~al.}(2016)Qiao, Zhang, and Jiang]{Qiao2016}
Qiao,~Z.; Zhang,~K.; Jiang,~J. {In silico screening of 4764 computation-ready, experimental metal-organic frameworks for CO$_{2}$ separation}. \emph{Journal of Materials Chemistry A} \textbf{2016}, \emph{4}, 2105--2114\relax
\mciteBstWouldAddEndPuncttrue
\mciteSetBstMidEndSepPunct{\mcitedefaultmidpunct}
{\mcitedefaultendpunct}{\mcitedefaultseppunct}\relax
\EndOfBibitem
\bibitem[Islamov \latin{et~al.}(2023)Islamov, Babaei, Anderson, Sezginel, Long, McGaughey, Gomez-Gualdron, and Wilmer]{Islamov2023}
Islamov,~M.; Babaei,~H.; Anderson,~R.; Sezginel,~K.~B.; Long,~J.~R.; McGaughey,~A.~J.; Gomez-Gualdron,~D.~A.; Wilmer,~C.~E. {High-throughput screening of hypothetical metal-organic frameworks for thermal conductivity}. \emph{npj Computational Materials} \textbf{2023}, \emph{9}\relax
\mciteBstWouldAddEndPuncttrue
\mciteSetBstMidEndSepPunct{\mcitedefaultmidpunct}
{\mcitedefaultendpunct}{\mcitedefaultseppunct}\relax
\EndOfBibitem
\bibitem[Schmidt \latin{et~al.}(2015)Schmidt, Yu, and McDaniel]{Schmidt2015}
Schmidt,~J.~R.; Yu,~K.; McDaniel,~J.~G. {Transferable Next-Generation Force Fields from Simple Liquids to Complex Materials}. \emph{Accounts of Chemical Research} \textbf{2015}, \emph{48}, 548--556\relax
\mciteBstWouldAddEndPuncttrue
\mciteSetBstMidEndSepPunct{\mcitedefaultmidpunct}
{\mcitedefaultendpunct}{\mcitedefaultseppunct}\relax
\EndOfBibitem
\bibitem[Harrison \latin{et~al.}(2018)Harrison, Schall, Maskey, Mikulski, Knippenberg, and Morrow]{Harrison2018}
Harrison,~J.~A.; Schall,~J.~D.; Maskey,~S.; Mikulski,~P.~T.; Knippenberg,~M.~T.; Morrow,~B.~H. {Review of force fields and intermolecular potentials used in atomistic computational materials research}. \textbf{2018}, \emph{5}\relax
\mciteBstWouldAddEndPuncttrue
\mciteSetBstMidEndSepPunct{\mcitedefaultmidpunct}
{\mcitedefaultendpunct}{\mcitedefaultseppunct}\relax
\EndOfBibitem
\bibitem[Bureekaew \latin{et~al.}(2013)Bureekaew, Amirjalayer, Tafipolsky, Spickermann, Roy, and Schmid]{Bureekaew2013}
Bureekaew,~S.; Amirjalayer,~S.; Tafipolsky,~M.; Spickermann,~C.; Roy,~T.~K.; Schmid,~R. {MOF-FF - A flexible first-principles derived force field for metal-organic frameworks}. \emph{Physica Status Solidi (B) Basic Research} \textbf{2013}, \emph{250}, 1128--1141\relax
\mciteBstWouldAddEndPuncttrue
\mciteSetBstMidEndSepPunct{\mcitedefaultmidpunct}
{\mcitedefaultendpunct}{\mcitedefaultseppunct}\relax
\EndOfBibitem
\bibitem[Heinen and Dubbeldam(2018)Heinen, and Dubbeldam]{Heinen2018}
Heinen,~J.; Dubbeldam,~D. On flexible force fields for metal–organic frameworks: Recent developments and future prospects. \emph{WIREs Computational Molecular Science} \textbf{2018}, \emph{8}, e1363\relax
\mciteBstWouldAddEndPuncttrue
\mciteSetBstMidEndSepPunct{\mcitedefaultmidpunct}
{\mcitedefaultendpunct}{\mcitedefaultseppunct}\relax
\EndOfBibitem
\bibitem[Unke \latin{et~al.}(2021)Unke, Chmiela, Sauceda, Gastegger, Poltavsky, Sch{\"{u}}tt, Tkatchenko, and M{\"{u}}ller]{Unke2021}
Unke,~O.~T.; Chmiela,~S.; Sauceda,~H.~E.; Gastegger,~M.; Poltavsky,~I.; Sch{\"{u}}tt,~K.~T.; Tkatchenko,~A.; M{\"{u}}ller,~K.~R. {Machine Learning Force Fields}. \textbf{2021}, \emph{121}, 10142--10186\relax
\mciteBstWouldAddEndPuncttrue
\mciteSetBstMidEndSepPunct{\mcitedefaultmidpunct}
{\mcitedefaultendpunct}{\mcitedefaultseppunct}\relax
\EndOfBibitem
\bibitem[Ko and Ong(2023)Ko, and Ong]{Ko2023}
Ko,~T.~W.; Ong,~S.~P. {Recent advances and outstanding challenges for machine learning interatomic potentials}. \emph{Nature Computational Science} \textbf{2023}, \emph{3}, 998--1000\relax
\mciteBstWouldAddEndPuncttrue
\mciteSetBstMidEndSepPunct{\mcitedefaultmidpunct}
{\mcitedefaultendpunct}{\mcitedefaultseppunct}\relax
\EndOfBibitem
\bibitem[Chen \latin{et~al.}(2019)Chen, Ye, Zuo, Zheng, and Ong]{MEGNet}
Chen,~C.; Ye,~W.; Zuo,~Y.; Zheng,~C.; Ong,~S.~P. {Graph Networks as a Universal Machine Learning Framework for Molecules and Crystals}. \emph{Chemistry of Materials} \textbf{2019}, \emph{31}, 3564--3572\relax
\mciteBstWouldAddEndPuncttrue
\mciteSetBstMidEndSepPunct{\mcitedefaultmidpunct}
{\mcitedefaultendpunct}{\mcitedefaultseppunct}\relax
\EndOfBibitem
\bibitem[Chen and Ong(2022)Chen, and Ong]{M3GNet}
Chen,~C.; Ong,~S.~P. {A universal graph deep learning interatomic potential for the periodic table}. \emph{Nature Computational Science} \textbf{2022}, \emph{2}, 718--728\relax
\mciteBstWouldAddEndPuncttrue
\mciteSetBstMidEndSepPunct{\mcitedefaultmidpunct}
{\mcitedefaultendpunct}{\mcitedefaultseppunct}\relax
\EndOfBibitem
\bibitem[Deng \latin{et~al.}(2023)Deng, Zhong, Jun, Riebesell, Han, Bartel, and Ceder]{Deng2023}
Deng,~B.; Zhong,~P.; Jun,~K.~J.; Riebesell,~J.; Han,~K.; Bartel,~C.~J.; Ceder,~G. {CHGNet as a pretrained universal neural network potential for charge-informed atomistic modelling}. \emph{Nature Machine Intelligence} \textbf{2023}, \emph{5}, 1031--1041\relax
\mciteBstWouldAddEndPuncttrue
\mciteSetBstMidEndSepPunct{\mcitedefaultmidpunct}
{\mcitedefaultendpunct}{\mcitedefaultseppunct}\relax
\EndOfBibitem
\bibitem[Batatia \latin{et~al.}(2022)Batatia, Kov{\'{a}}cs, Simm, Ortner, and Cs{\'{a}}nyi]{MACE_arch}
Batatia,~I.; Kov{\'{a}}cs,~D.~P.; Simm,~G. N.~C.; Ortner,~C.; Cs{\'{a}}nyi,~G. {MACE: Higher Order Equivariant Message Passing Neural Networks for Fast and Accurate Force Fields}. 2022; \url{http://arxiv.org/abs/2206.07697}\relax
\mciteBstWouldAddEndPuncttrue
\mciteSetBstMidEndSepPunct{\mcitedefaultmidpunct}
{\mcitedefaultendpunct}{\mcitedefaultseppunct}\relax
\EndOfBibitem
\bibitem[Drautz(2019)]{Drautz2019}
Drautz,~R. {Atomic cluster expansion for accurate and transferable interatomic potentials}. \emph{Physical Review B} \textbf{2019}, \emph{99}\relax
\mciteBstWouldAddEndPuncttrue
\mciteSetBstMidEndSepPunct{\mcitedefaultmidpunct}
{\mcitedefaultendpunct}{\mcitedefaultseppunct}\relax
\EndOfBibitem
\bibitem[Batatia \latin{et~al.}(2023)Batatia, Benner, Chiang, Elena, Kov{\'{a}}cs, Riebesell, Advincula, Asta, Baldwin, Bernstein, Bhowmik, Blau, C{\u{a}}rare, Darby, De, Della~Pia, Deringer, Elijo{\v{s}}ius, El-Machachi, Fako, Ferrari, Genreith-Schriever, George, Goodall, Grey, Han, Handley, Heenen, Hermansson, Holm, Jaafar, Hofmann, Jakob, Jung, Kapil, Kaplan, Karimitari, Kroupa, Kullgren, Kuner, Kuryla, Liepuoniute, Margraf, Magd{\u{a}}u, Michaelides, Moore, Naik, Niblett, Norwood, O'Neill, Ortner, Persson, Reuter, Rosen, Schaaf, Schran, Sivonxay, Stenczel, Svahn, Sutton, van~der Oord, Varga-Umbrich, Vegge, Vondr{\'{a}}k, Wang, Witt, Zills, and Cs{\'{a}}nyi]{MACE-MP-0}
Batatia,~I. \latin{et~al.}  {A foundation model for atomistic materials chemistry}. 2023; \url{http://arxiv.org/abs/2401.00096}\relax
\mciteBstWouldAddEndPuncttrue
\mciteSetBstMidEndSepPunct{\mcitedefaultmidpunct}
{\mcitedefaultendpunct}{\mcitedefaultseppunct}\relax
\EndOfBibitem
\bibitem[Rosen \latin{et~al.}(2021)Rosen, Iyer, Ray, Yao, Aspuru-Guzik, Gagliardi, Notestein, and Snurr]{Rosen2021}
Rosen,~A.~S.; Iyer,~S.~M.; Ray,~D.; Yao,~Z.; Aspuru-Guzik,~A.; Gagliardi,~L.; Notestein,~J.~M.; Snurr,~R.~Q. {Machine learning the quantum-chemical properties of metal–organic frameworks for accelerated materials discovery}. \emph{Matter} \textbf{2021}, \emph{4}, 1578--1597\relax
\mciteBstWouldAddEndPuncttrue
\mciteSetBstMidEndSepPunct{\mcitedefaultmidpunct}
{\mcitedefaultendpunct}{\mcitedefaultseppunct}\relax
\EndOfBibitem
\bibitem[Jain \latin{et~al.}(2013)Jain, Ong, Hautier, Chen, Richards, Dacek, Cholia, Gunter, Skinner, Ceder, and Persson]{Jain2013_MP}
Jain,~A.; Ong,~S.~P.; Hautier,~G.; Chen,~W.; Richards,~W.~D.; Dacek,~S.; Cholia,~S.; Gunter,~D.; Skinner,~D.; Ceder,~G.; Persson,~K.~A. {Commentary: The materials project: A materials genome approach to accelerating materials innovation}. \textbf{2013}, \emph{1}\relax
\mciteBstWouldAddEndPuncttrue
\mciteSetBstMidEndSepPunct{\mcitedefaultmidpunct}
{\mcitedefaultendpunct}{\mcitedefaultseppunct}\relax
\EndOfBibitem
\bibitem[Ghahremanpour \latin{et~al.}(2018)Ghahremanpour, Van~Maaren, and Van Der~Spoel]{alexandriaDatabase}
Ghahremanpour,~M.~M.; Van~Maaren,~P.~J.; Van Der~Spoel,~D. {Data Descriptor: The Alexandria library, a quantum-chemical database of molecular properties for force field development}. \emph{Scientific Data} \textbf{2018}, \emph{5}\relax
\mciteBstWouldAddEndPuncttrue
\mciteSetBstMidEndSepPunct{\mcitedefaultmidpunct}
{\mcitedefaultendpunct}{\mcitedefaultseppunct}\relax
\EndOfBibitem
\bibitem[Barroso-Luque \latin{et~al.}(2024)Barroso-Luque, Shuaibi, Fu, Wood, Dzamba, Gao, Rizvi, Zitnick, and Ulissi]{OMAT24}
Barroso-Luque,~L.; Shuaibi,~M.; Fu,~X.; Wood,~B.~M.; Dzamba,~M.; Gao,~M.; Rizvi,~A.; Zitnick,~C.~L.; Ulissi,~Z.~W. {Open Materials 2024 (OMat24) Inorganic Materials Dataset and Models}. 2024; \url{http://arxiv.org/abs/2410.12771}\relax
\mciteBstWouldAddEndPuncttrue
\mciteSetBstMidEndSepPunct{\mcitedefaultmidpunct}
{\mcitedefaultendpunct}{\mcitedefaultseppunct}\relax
\EndOfBibitem
\bibitem[Riebesell \latin{et~al.}(2023)Riebesell, Goodall, Benner, Chiang, Deng, Lee, Jain, and Persson]{Riebesell2023}
Riebesell,~J.; Goodall,~R. E.~A.; Benner,~P.; Chiang,~Y.; Deng,~B.; Lee,~A.~A.; Jain,~A.; Persson,~K.~A. {Matbench Discovery -- A framework to evaluate machine learning crystal stability predictions}. 2023; \url{http://arxiv.org/abs/2308.14920}\relax
\mciteBstWouldAddEndPuncttrue
\mciteSetBstMidEndSepPunct{\mcitedefaultmidpunct}
{\mcitedefaultendpunct}{\mcitedefaultseppunct}\relax
\EndOfBibitem
\bibitem[Fu \latin{et~al.}(2025)Fu, Wood, Barroso-Luque, Levine, Gao, Dzamba, and Zitnick]{eSEN}
Fu,~X.; Wood,~B.~M.; Barroso-Luque,~L.; Levine,~D.~S.; Gao,~M.; Dzamba,~M.; Zitnick,~C.~L. {Learning Smooth and Expressive Interatomic Potentials for Physical Property Prediction}. 2025; \url{http://arxiv.org/abs/2502.12147}\relax
\mciteBstWouldAddEndPuncttrue
\mciteSetBstMidEndSepPunct{\mcitedefaultmidpunct}
{\mcitedefaultendpunct}{\mcitedefaultseppunct}\relax
\EndOfBibitem
\bibitem[Focassio \latin{et~al.}(2024)Focassio, Freitas, and Schleder]{Focassio2024}
Focassio,~B.; Freitas,~L. P.~M.; Schleder,~G.~R. Performance Assessment of Universal Machine Learning Interatomic Potentials: Challenges and Directions for Materials' Surfaces. \textit{ChemRxiv}, May 30, \textbf{2024}, ver. 2, 2024; \url{https://doi.org/10.48550/arXiv.2403.04217}\relax
\mciteBstWouldAddEndPuncttrue
\mciteSetBstMidEndSepPunct{\mcitedefaultmidpunct}
{\mcitedefaultendpunct}{\mcitedefaultseppunct}\relax
\EndOfBibitem
\bibitem[Zheng \latin{et~al.}(2023)Zheng, Oliveira, Neumann Barros~Ferreira, Steiner, Hamann, Gu, and Luan]{Zheng2023}
Zheng,~B.; Oliveira,~F.~L.; Neumann Barros~Ferreira,~R.; Steiner,~M.; Hamann,~H.; Gu,~G.~X.; Luan,~B. {Quantum Informed Machine-Learning Potentials for Molecular Dynamics Simulations of CO$_{2}$’s Chemisorption and Diffusion in Mg-MOF-74}. \emph{ACS Nano} \textbf{2023}, \emph{17}, 5579--5587\relax
\mciteBstWouldAddEndPuncttrue
\mciteSetBstMidEndSepPunct{\mcitedefaultmidpunct}
{\mcitedefaultendpunct}{\mcitedefaultseppunct}\relax
\EndOfBibitem
\bibitem[Zheng \latin{et~al.}(2024)Zheng, Gu, Santos, Neumann Barros~Ferreira, Steiner, and Luan]{Zheng2024}
Zheng,~B.; Gu,~G.~X.; Santos,~C.~d.; Neumann Barros~Ferreira,~R.; Steiner,~M.; Luan,~B. {Simulating CO$_2$ diffusivity in rigid and flexible Mg-MOF-74 with machine-learning force fields}. \emph{APL Machine Learning} \textbf{2024}, \emph{2}, 026115\relax
\mciteBstWouldAddEndPuncttrue
\mciteSetBstMidEndSepPunct{\mcitedefaultmidpunct}
{\mcitedefaultendpunct}{\mcitedefaultseppunct}\relax
\EndOfBibitem
\bibitem[Liao and Smidt(2022)Liao, and Smidt]{Liao2022_equi}
Liao,~Y.-L.; Smidt,~T. {Equiformer: Equivariant Graph Attention Transformer for 3D Atomistic Graphs}. 2022; \url{http://arxiv.org/abs/2206.11990}\relax
\mciteBstWouldAddEndPuncttrue
\mciteSetBstMidEndSepPunct{\mcitedefaultmidpunct}
{\mcitedefaultendpunct}{\mcitedefaultseppunct}\relax
\EndOfBibitem
\bibitem[Chanussot \latin{et~al.}(2021)Chanussot, Das, Goyal, Lavril, Shuaibi, Riviere, Tran, Heras-Domingo, Ho, Hu, Palizhati, Sriram, Wood, Yoon, Parikh, Zitnick, and Ulissi]{OC20}
Chanussot,~L. \latin{et~al.}  {Open Catalyst 2020 (OC20) Dataset and Community Challenges}. \emph{ACS Catalysis} \textbf{2021}, \emph{11}, 6059--6072\relax
\mciteBstWouldAddEndPuncttrue
\mciteSetBstMidEndSepPunct{\mcitedefaultmidpunct}
{\mcitedefaultendpunct}{\mcitedefaultseppunct}\relax
\EndOfBibitem
\bibitem[Tran \latin{et~al.}(2023)Tran, Lan, Shuaibi, Wood, Goyal, Das, Heras-Domingo, Kolluru, Rizvi, Shoghi, Sriram, Therrien, Abed, Voznyy, Sargent, Ulissi, and Zitnick]{OC22}
Tran,~R. \latin{et~al.}  {The Open Catalyst 2022 (OC22) Dataset and Challenges for Oxide Electrocatalysts}. \emph{ACS Catalysis} \textbf{2023}, \emph{13}, 3066--3084\relax
\mciteBstWouldAddEndPuncttrue
\mciteSetBstMidEndSepPunct{\mcitedefaultmidpunct}
{\mcitedefaultendpunct}{\mcitedefaultseppunct}\relax
\EndOfBibitem
\bibitem[Findley and Sholl(2021)Findley, and Sholl]{Findley2021}
Findley,~J.~M.; Sholl,~D.~S. Computational Screening of MOFs and Zeolites for Direct Air Capture of Carbon Dioxide under Humid Conditions. \emph{The Journal of Physical Chemistry C} \textbf{2021}, \emph{125}, 24630--24639\relax
\mciteBstWouldAddEndPuncttrue
\mciteSetBstMidEndSepPunct{\mcitedefaultmidpunct}
{\mcitedefaultendpunct}{\mcitedefaultseppunct}\relax
\EndOfBibitem
\bibitem[Addicoat \latin{et~al.}(2014)Addicoat, Vankova, Akter, and Heine]{UFF4MOF1}
Addicoat,~M.~A.; Vankova,~N.; Akter,~I.~F.; Heine,~T. {Extension of the universal force field to metal-organic frameworks}. \emph{Journal of Chemical Theory and Computation} \textbf{2014}, \emph{10}, 880--891\relax
\mciteBstWouldAddEndPuncttrue
\mciteSetBstMidEndSepPunct{\mcitedefaultmidpunct}
{\mcitedefaultendpunct}{\mcitedefaultseppunct}\relax
\EndOfBibitem
\bibitem[Coupry \latin{et~al.}(2016)Coupry, Addicoat, and Heine]{UFF4MOF2}
Coupry,~D.~E.; Addicoat,~M.~A.; Heine,~T. {Extension of the Universal Force Field for Metal-Organic Frameworks}. \emph{Journal of Chemical Theory and Computation} \textbf{2016}, \emph{12}, 5215--5225\relax
\mciteBstWouldAddEndPuncttrue
\mciteSetBstMidEndSepPunct{\mcitedefaultmidpunct}
{\mcitedefaultendpunct}{\mcitedefaultseppunct}\relax
\EndOfBibitem
\bibitem[Perdew \latin{et~al.}(1996)Perdew, Burke, and Ernzerhof]{PBE}
Perdew,~J.~P.; Burke,~K.; Ernzerhof,~M. {Generalized Gradient Approximation Made Simple}. \emph{Physical Review Letters} \textbf{1996}, \emph{77}, 3865--3868\relax
\mciteBstWouldAddEndPuncttrue
\mciteSetBstMidEndSepPunct{\mcitedefaultmidpunct}
{\mcitedefaultendpunct}{\mcitedefaultseppunct}\relax
\EndOfBibitem
\bibitem[Grimme \latin{et~al.}(2010)Grimme, Antony, Ehrlich, and Krieg]{Grimme_DFTD}
Grimme,~S.; Antony,~J.; Ehrlich,~S.; Krieg,~H. {A consistent and accurate ab initio parametrization of density functional dispersion correction (DFT-D) for the 94 elements H-Pu}. \emph{Journal of Chemical Physics} \textbf{2010}, \emph{132}\relax
\mciteBstWouldAddEndPuncttrue
\mciteSetBstMidEndSepPunct{\mcitedefaultmidpunct}
{\mcitedefaultendpunct}{\mcitedefaultseppunct}\relax
\EndOfBibitem
\bibitem[Kresse and Furthm\"{u}ller(1996)Kresse, and Furthm\"{u}ller]{VASP}
Kresse,~G.; Furthm\"{u}ller,~J. {Efficiency of ab-initio total energy calculations for metals and semiconductors using a plane-wave basis set}. \textbf{1996}, \emph{6}, 15--50\relax
\mciteBstWouldAddEndPuncttrue
\mciteSetBstMidEndSepPunct{\mcitedefaultmidpunct}
{\mcitedefaultendpunct}{\mcitedefaultseppunct}\relax
\EndOfBibitem
\bibitem[Manz and Sholl(2010)Manz, and Sholl]{DDEC}
Manz,~T.~A.; Sholl,~D.~S. {Chemically meaningful atomic charges that reproduce the electrostatic potential in periodic and nonperiodic materials}. \emph{Journal of Chemical Theory and Computation} \textbf{2010}, \emph{6}, 2455--2468\relax
\mciteBstWouldAddEndPuncttrue
\mciteSetBstMidEndSepPunct{\mcitedefaultmidpunct}
{\mcitedefaultendpunct}{\mcitedefaultseppunct}\relax
\EndOfBibitem
\bibitem[Potoff and Siepmann(2001)Potoff, and Siepmann]{Potoff_CO2}
Potoff,~J.~J.; Siepmann,~J.~I. {Vapor-liquid equilibria of mixtures containing alkanes, carbon dioxide, and nitrogen}. \emph{AIChE Journal} \textbf{2001}, \emph{47}, 1676--1682\relax
\mciteBstWouldAddEndPuncttrue
\mciteSetBstMidEndSepPunct{\mcitedefaultmidpunct}
{\mcitedefaultendpunct}{\mcitedefaultseppunct}\relax
\EndOfBibitem
\bibitem[Eggimann \latin{et~al.}(2014)Eggimann, Sunnarborg, Stern, Bliss, and Siepmann]{TraPPE}
Eggimann,~B.~L.; Sunnarborg,~A.~J.; Stern,~H.~D.; Bliss,~A.~P.; Siepmann,~J.~I. {An online parameter and property database for the TraPPE force field}. \emph{Molecular Simulation} \textbf{2014}, \emph{40}, 101--105\relax
\mciteBstWouldAddEndPuncttrue
\mciteSetBstMidEndSepPunct{\mcitedefaultmidpunct}
{\mcitedefaultendpunct}{\mcitedefaultseppunct}\relax
\EndOfBibitem
\bibitem[Berendsen \latin{et~al.}(1987)Berendsen, Grigera, and Straatsma]{SPCE}
Berendsen,~H. J.~C.; Grigera,~J.~R.; Straatsma,~T.~P. {The Missing Term in Effective Pair Potentials1}. \emph{J. Phys. Chem} \textbf{1987}, \emph{91}, 6269--6271\relax
\mciteBstWouldAddEndPuncttrue
\mciteSetBstMidEndSepPunct{\mcitedefaultmidpunct}
{\mcitedefaultendpunct}{\mcitedefaultseppunct}\relax
\EndOfBibitem
\bibitem[Jorgensen \latin{et~al.}(1983)Jorgensen, Chandrasekhar, Madura, Impey, and Klein]{TIP4P}
Jorgensen,~W.~L.; Chandrasekhar,~J.; Madura,~J.~D.; Impey,~R.~W.; Klein,~M.~L. {Comparison of simple potential functions for simulating liquid water}. \emph{The Journal of Chemical Physics} \textbf{1983}, \emph{79}, 926--935\relax
\mciteBstWouldAddEndPuncttrue
\mciteSetBstMidEndSepPunct{\mcitedefaultmidpunct}
{\mcitedefaultendpunct}{\mcitedefaultseppunct}\relax
\EndOfBibitem
\bibitem[Chen \latin{et~al.}(2000)Chen, Xing, and Siepmann]{Chen2000}
Chen,~B.; Xing,~J.; Siepmann,~J.~I. {Development of Polarizable Water Force Fields for Phase Equilibrium Calculations}. \emph{Journal of Physical Chemistry B} \textbf{2000}, \emph{104}, 2391--2401\relax
\mciteBstWouldAddEndPuncttrue
\mciteSetBstMidEndSepPunct{\mcitedefaultmidpunct}
{\mcitedefaultendpunct}{\mcitedefaultseppunct}\relax
\EndOfBibitem
\bibitem[Zielkiewicz(2005)]{Zielkiewicz2005}
Zielkiewicz,~J. {Structural properties of water: Comparison of the SPC, SPCE, TIP4P, and TIP5P models of water}. \emph{Journal of Chemical Physics} \textbf{2005}, \emph{123}\relax
\mciteBstWouldAddEndPuncttrue
\mciteSetBstMidEndSepPunct{\mcitedefaultmidpunct}
{\mcitedefaultendpunct}{\mcitedefaultseppunct}\relax
\EndOfBibitem
\bibitem[Lorentz(1881)]{Lorentz1881}
Lorentz,~H.~A. Ueber die Anwendung des Satzes vom Virial in der kinetischen Theorie der Gase. \emph{Annalen der Physik} \textbf{1881}, \emph{248}, 127--136\relax
\mciteBstWouldAddEndPuncttrue
\mciteSetBstMidEndSepPunct{\mcitedefaultmidpunct}
{\mcitedefaultendpunct}{\mcitedefaultseppunct}\relax
\EndOfBibitem
\bibitem[Berthelot(1889)]{Berthelot1889}
Berthelot,~D. Sur le m\'{e} lange des gaz. \emph{C. R. Acad. Sci. Paris} \textbf{1889}, \emph{126}\relax
\mciteBstWouldAddEndPuncttrue
\mciteSetBstMidEndSepPunct{\mcitedefaultmidpunct}
{\mcitedefaultendpunct}{\mcitedefaultseppunct}\relax
\EndOfBibitem
\bibitem[Ewald(1921)]{Ewald1921}
Ewald,~P.~P. Die Berechnung optischer und elektrostatischer Gitterpotentiale. \emph{Annalen der Physik} \textbf{1921}, \emph{369}, 253--287\relax
\mciteBstWouldAddEndPuncttrue
\mciteSetBstMidEndSepPunct{\mcitedefaultmidpunct}
{\mcitedefaultendpunct}{\mcitedefaultseppunct}\relax
\EndOfBibitem
\bibitem[Cleeton \latin{et~al.}(2023)Cleeton, de~Oliveira, Neumann, Farmahini, Luan, Steiner, and Sarkisov]{Cleeton2023}
Cleeton,~C.; de~Oliveira,~F.~L.; Neumann,~R.~F.; Farmahini,~A.~H.; Luan,~B.; Steiner,~M.; Sarkisov,~L. {A process-level perspective of the impact of molecular force fields on the computational screening of MOFs for carbon capture}. \emph{Energy and Environmental Science} \textbf{2023}, \emph{16}, 3899--3918\relax
\mciteBstWouldAddEndPuncttrue
\mciteSetBstMidEndSepPunct{\mcitedefaultmidpunct}
{\mcitedefaultendpunct}{\mcitedefaultseppunct}\relax
\EndOfBibitem
\bibitem[Thompson \latin{et~al.}(2022)Thompson, Aktulga, Berger, Bolintineanu, Brown, Crozier, in~'t Veld, Kohlmeyer, Moore, Nguyen, Shan, Stevens, Tranchida, Trott, and Plimpton]{LAMMPS}
Thompson,~A.~P.; Aktulga,~H.~M.; Berger,~R.; Bolintineanu,~D.~S.; Brown,~W.~M.; Crozier,~P.~S.; in~'t Veld,~P.~J.; Kohlmeyer,~A.; Moore,~S.~G.; Nguyen,~T.~D.; Shan,~R.; Stevens,~M.~J.; Tranchida,~J.; Trott,~C.; Plimpton,~S.~J. {LAMMPS - a flexible simulation tool for particle-based materials modeling at the atomic, meso, and continuum scales}. \emph{Computer Physics Communications} \textbf{2022}, \emph{271}\relax
\mciteBstWouldAddEndPuncttrue
\mciteSetBstMidEndSepPunct{\mcitedefaultmidpunct}
{\mcitedefaultendpunct}{\mcitedefaultseppunct}\relax
\EndOfBibitem
\bibitem[Rouge \latin{et~al.}(1969)Rouge, Polak, and Ribi{\`e}re]{PolakRibiere}
Rouge,~S.; Polak,~E.; Ribi{\`e}re,~G. Note sur la convergence de m{\'e}thodes de directions conjugu{\'e}es. \emph{R.I.R.O.} \textbf{1969}, 35--43\relax
\mciteBstWouldAddEndPuncttrue
\mciteSetBstMidEndSepPunct{\mcitedefaultmidpunct}
{\mcitedefaultendpunct}{\mcitedefaultseppunct}\relax
\EndOfBibitem
\bibitem[Liao and Smidt(2022)Liao, and Smidt]{Equiformer_arch}
Liao,~Y.-L.; Smidt,~T. {Equiformer: Equivariant Graph Attention Transformer for 3D Atomistic Graphs}. 2022; \url{http://arxiv.org/abs/2206.11990}\relax
\mciteBstWouldAddEndPuncttrue
\mciteSetBstMidEndSepPunct{\mcitedefaultmidpunct}
{\mcitedefaultendpunct}{\mcitedefaultseppunct}\relax
\EndOfBibitem
\bibitem[Hjorth~Larsen \latin{et~al.}(2017)Hjorth~Larsen, J{\O}rgen~Mortensen, Blomqvist, Castelli, Christensen, Du{\l}ak, Friis, Groves, Hammer, Hargus, Hermes, Jennings, Bjerre~Jensen, Kermode, Kitchin, Leonhard~Kolsbjerg, Kubal, Kaasbjerg, Lysgaard, Bergmann~Maronsson, Maxson, Olsen, Pastewka, Peterson, Rostgaard, Schi{\O}tz, Sch{\"{u}}tt, Strange, Thygesen, Vegge, Vilhelmsen, Walter, Zeng, and Jacobsen]{ASE}
Hjorth~Larsen,~A. \latin{et~al.}  {The atomic simulation environment - A Python library for working with atoms}. \emph{Journal of Physics Condensed Matter} \textbf{2017}, \emph{29}\relax
\mciteBstWouldAddEndPuncttrue
\mciteSetBstMidEndSepPunct{\mcitedefaultmidpunct}
{\mcitedefaultendpunct}{\mcitedefaultseppunct}\relax
\EndOfBibitem
\bibitem[Chung \latin{et~al.}(2014)Chung, Camp, Haranczyk, Sikora, Bury, Krungleviciute, Yildirim, Farha, Sholl, and Snurr]{CoRE_MOF}
Chung,~Y.~G.; Camp,~J.; Haranczyk,~M.; Sikora,~B.~J.; Bury,~W.; Krungleviciute,~V.; Yildirim,~T.; Farha,~O.~K.; Sholl,~D.~S.; Snurr,~R.~Q. {Computation-ready, experimental metal-organic frameworks: A tool to enable high-throughput screening of nanoporous crystals}. \emph{Chemistry of Materials} \textbf{2014}, \emph{26}, 6185--6192\relax
\mciteBstWouldAddEndPuncttrue
\mciteSetBstMidEndSepPunct{\mcitedefaultmidpunct}
{\mcitedefaultendpunct}{\mcitedefaultseppunct}\relax
\EndOfBibitem
\bibitem[Chung \latin{et~al.}(2019)Chung, Haldoupis, Bucior, Haranczyk, Lee, Zhang, Vogiatzis, Milisavljevic, Ling, Camp, Slater, Siepmann, Sholl, and Snurr]{CoRE2019}
Chung,~Y.~G.; Haldoupis,~E.; Bucior,~B.~J.; Haranczyk,~M.; Lee,~S.; Zhang,~H.; Vogiatzis,~K.~D.; Milisavljevic,~M.; Ling,~S.; Camp,~J.~S.; Slater,~B.; Siepmann,~J.~I.; Sholl,~D.~S.; Snurr,~R.~Q. {Advances, Updates, and Analytics for the Computation-Ready, Experimental Metal-Organic Framework Database: CoRE MOF 2019}. \emph{Journal of Chemical and Engineering Data} \textbf{2019}, \emph{64}, 5985--5998\relax
\mciteBstWouldAddEndPuncttrue
\mciteSetBstMidEndSepPunct{\mcitedefaultmidpunct}
{\mcitedefaultendpunct}{\mcitedefaultseppunct}\relax
\EndOfBibitem
\bibitem[Chen and Manz(2020)Chen, and Manz]{ChenManz2020}
Chen,~T.; Manz,~T.~A. {Identifying misbonded atoms in the 2019 CoRE metal-organic framework database}. \emph{RSC Advances} \textbf{2020}, \emph{10}, 26944--26951\relax
\mciteBstWouldAddEndPuncttrue
\mciteSetBstMidEndSepPunct{\mcitedefaultmidpunct}
{\mcitedefaultendpunct}{\mcitedefaultseppunct}\relax
\EndOfBibitem
\bibitem[White \latin{et~al.}(2024)White, Gibaldi, Burner, Mayo, and Woo]{White2024}
White,~A.; Gibaldi,~M.; Burner,~J.; Mayo,~R.~A.; Woo,~T. {Alarming structural error rates in MOF databases used in data driven workflows identified via a novel metal oxidation state-based method}. 2024; \url{https://chemrxiv.org/engage/chemrxiv/article-details/6706b96312ff75c3a1fb0365}\relax
\mciteBstWouldAddEndPuncttrue
\mciteSetBstMidEndSepPunct{\mcitedefaultmidpunct}
{\mcitedefaultendpunct}{\mcitedefaultseppunct}\relax
\EndOfBibitem
\bibitem[Gibaldi \latin{et~al.}(2025)Gibaldi, Kapeliukha, White, Luo, Mayo, Burner, and Woo]{Gibaldi2025}
Gibaldi,~M.; Kapeliukha,~A.; White,~A.; Luo,~J.; Mayo,~R.~A.; Burner,~J.; Woo,~T.~K. {MOSAEC-DB: a comprehensive database of experimental metal-organic frameworks with verified chemical accuracy suitable for molecular simulations}. \emph{Chemical Science} \textbf{2025}, \emph{16}, 4085--4100\relax
\mciteBstWouldAddEndPuncttrue
\mciteSetBstMidEndSepPunct{\mcitedefaultmidpunct}
{\mcitedefaultendpunct}{\mcitedefaultseppunct}\relax
\EndOfBibitem
\bibitem[Zhao \latin{et~al.}(2025)Zhao, Brabson, Chheda, Huang, Kim, Liu, Mochida, Pham, {Prerna}, Terrones, Yoon, Zoubritzky, Coudert, Haranczyk, Kulik, Moosavi, Sholl, Siepmann, Snurr, and Chung]{CoRE2025}
Zhao,~G. \latin{et~al.}  {CoRE MOF DB: A curated experimental metal-organic framework database with machine-learned properties for integrated material-process screening}. \emph{Matter} \textbf{2025}, 102140\relax
\mciteBstWouldAddEndPuncttrue
\mciteSetBstMidEndSepPunct{\mcitedefaultmidpunct}
{\mcitedefaultendpunct}{\mcitedefaultseppunct}\relax
\EndOfBibitem
\bibitem[Jin \latin{et~al.}(2025)Jin, Jablonka, Moubarak, Li, and Smit]{MOFChecker}
Jin,~X.; Jablonka,~K.; Moubarak,~E.; Li,~Y.; Smit,~B. {MOFChecker: An algorithm for Validating and Correcting Metal-Organic Framework (MOF) Structures}. 2025; \url{https://chemrxiv.org/engage/chemrxiv/article-details/67aa2e72fa469535b9c2dcfd}\relax
\mciteBstWouldAddEndPuncttrue
\mciteSetBstMidEndSepPunct{\mcitedefaultmidpunct}
{\mcitedefaultendpunct}{\mcitedefaultseppunct}\relax
\EndOfBibitem
\bibitem[Wilmer \latin{et~al.}(2012)Wilmer, Kim, and Snurr]{Wilmer2012}
Wilmer,~C.~E.; Kim,~K.~C.; Snurr,~R.~Q. {An extended charge equilibration method}. \emph{Journal of Physical Chemistry Letters} \textbf{2012}, \emph{3}, 2506--2511\relax
\mciteBstWouldAddEndPuncttrue
\mciteSetBstMidEndSepPunct{\mcitedefaultmidpunct}
{\mcitedefaultendpunct}{\mcitedefaultseppunct}\relax
\EndOfBibitem
\bibitem[Nazarian \latin{et~al.}(2016)Nazarian, Camp, and Sholl]{Nazarian2016}
Nazarian,~D.; Camp,~J.~S.; Sholl,~D.~S. {A Comprehensive Set of High-Quality Point Charges for Simulations of Metal-Organic Frameworks}. \emph{Chemistry of Materials} \textbf{2016}, \emph{28}, 785--793\relax
\mciteBstWouldAddEndPuncttrue
\mciteSetBstMidEndSepPunct{\mcitedefaultmidpunct}
{\mcitedefaultendpunct}{\mcitedefaultseppunct}\relax
\EndOfBibitem
\bibitem[Pedregosa \latin{et~al.}(2012)Pedregosa, Varoquaux, Gramfort, Michel, Thirion, Grisel, Blondel, M{\"{u}}ller, Nothman, Louppe, Prettenhofer, Weiss, Dubourg, Vanderplas, Passos, Cournapeau, Brucher, Perrot, and Duchesnay]{sklearn}
Pedregosa,~F. \latin{et~al.}  {Scikit-learn: Machine Learning in Python}. 2012; \url{http://arxiv.org/abs/1201.0490}\relax
\mciteBstWouldAddEndPuncttrue
\mciteSetBstMidEndSepPunct{\mcitedefaultmidpunct}
{\mcitedefaultendpunct}{\mcitedefaultseppunct}\relax
\EndOfBibitem
\end{mcitethebibliography}
\end{document}